\documentclass[journal]{IEEEtran}

\usepackage{cite}
\usepackage{multirow}
\usepackage[cmex10]{amsmath}
\usepackage{amssymb}
\usepackage{array}
\usepackage{url}
\usepackage{ctable}
\usepackage{afterpage}
\usepackage{booktabs}
\usepackage{footnote}
\usepackage{threeparttable}
\usepackage{colortbl}
\usepackage{float}
\usepackage{subfig}
\usepackage[normalem]{ulem}

\begin{document}
	
\title{Analysis and Design of Commutation-Based Circulator-Receivers for Integrated Full-Duplex Wireless}

\author{Negar~Reiskarimian,~\IEEEmembership{Student Member, IEEE,}
	Mahmood~Baraani~Dastjerdi,~\IEEEmembership{Student Member, IEEE,}
	Jin~Zhou,~\IEEEmembership{Member, IEEE,} 
	and~Harish~Krishnaswamy,~\IEEEmembership{Member, IEEE}
}
	
\markboth{Submitted to IEEE Journal of Solid-State Circuits}{}
\maketitle

\begin{abstract}

Previously, we presented a non-magnetic, non-reciprocal N-path-filter-based circulator-receiver (circ.-RX) architecture for full-duplex (FD) wireless which merges a commutation-based linear periodically-time-varying (LPTV) non-magnetic circulator with a down-converting mixer and directly provides the baseband (BB) receiver signals at its output, while suppressing the noise contribution of one set of the commutating switches. The architecture also incorporates an on-chip balance network to enhance the transmitter (TX)-receiver (RX) isolation. In this paper, we present a detailed analysis of the architecture, including a noise analysis and an analysis of the effect of the balance network. The analyses are verified by simulation and measurement results of a 65~nm CMOS 750~MHz  circulator-receiver prototype. The circulator-receiver can handle up to +8~dBm of TX power, with 8~dB noise figure (NF) and 40~dB average isolation over 20~MHz RF bandwidth (BW). In conjunction with digital self-interference (SI) and its third-order intermodulation (IM3) cancellation, the FD circ.-RX demonstrates 80~dB overall SI suppression for up to +8~dBm TX average output power. The claims are also verified through an FD demonstration where a -50~dBm weak desired received signal is recovered while transmitting a 0~dBm average-power OFDM-like TX signal.
\end{abstract}
\vspace{-12pt}

\section{Introduction}
\IEEEPARstart{T}{he} thousand-fold data capacity increase envisioned in the next generation of wireless communication networks or "5G" is expected to be delivered by technology candidates such as full-duplex wireless and massive multiple-input multiple-output (MIMO) wireless \cite{CommMag_5G_tech,CommMag_5G_tech_2}. FD aims to instantly double the link capacity in the physical layer by simultaneously transmitting and receiving at the same frequency, as well as providing other benefits in the higher layers such as better spectral efficiency, reducing network and feedback signaling delays, and resolving hidden-node problems to avoid collisions \cite{FullDuplex_Katti,FullDuplexInvitedPaper_RiceJSAC14,FD_survey,Flexicon_Comm_Mag}.

Achieving the $2\times$ throughput gain and the other benefits mentioned above is contingent upon overcoming several fundamental challenges associated with FD operation. The first is the tremendous amount of self-interference from the transmitter at its own receiver. An FD system handling $+8~$dBm of transmit power, with $20~$MHz signal bandwidth and $8~$dB NF budget, requires almost $100~$dB of self-interference cancellation (SIC) to cancel the TX leakage down to the receiver noise floor of $-93~$dBm. This can only be achieved through multi-domain SIC at the antenna interface \cite{NRK_NatComm16, NRK_JSSC2017,EB_Duplexer_Entesari_TMTT2016, imec_TMTT_2016,EB_Duplexer_Larson_TMTT2016, TD_JSSC2015,Tolga_AntennaSIC_IMS15}, RF/analog baseband \cite{NRK_JSSC2017,Rudell_ISSCC2017,Berkeley_RFIC_2017,Danilo_RFIC2017,TD_JSSC2015,CornellFullDuplex_JSSC15,Zhou_FDESIC_JSSC15,TwenteSICRX_JSSC2015,Zhou_NCSIC_JSSC14} and in digital \cite{FullDuplex_Katti,NRK_JSSC2017}.

\begin{figure}[!t]
	\includegraphics[keepaspectratio,width=1.05\linewidth]{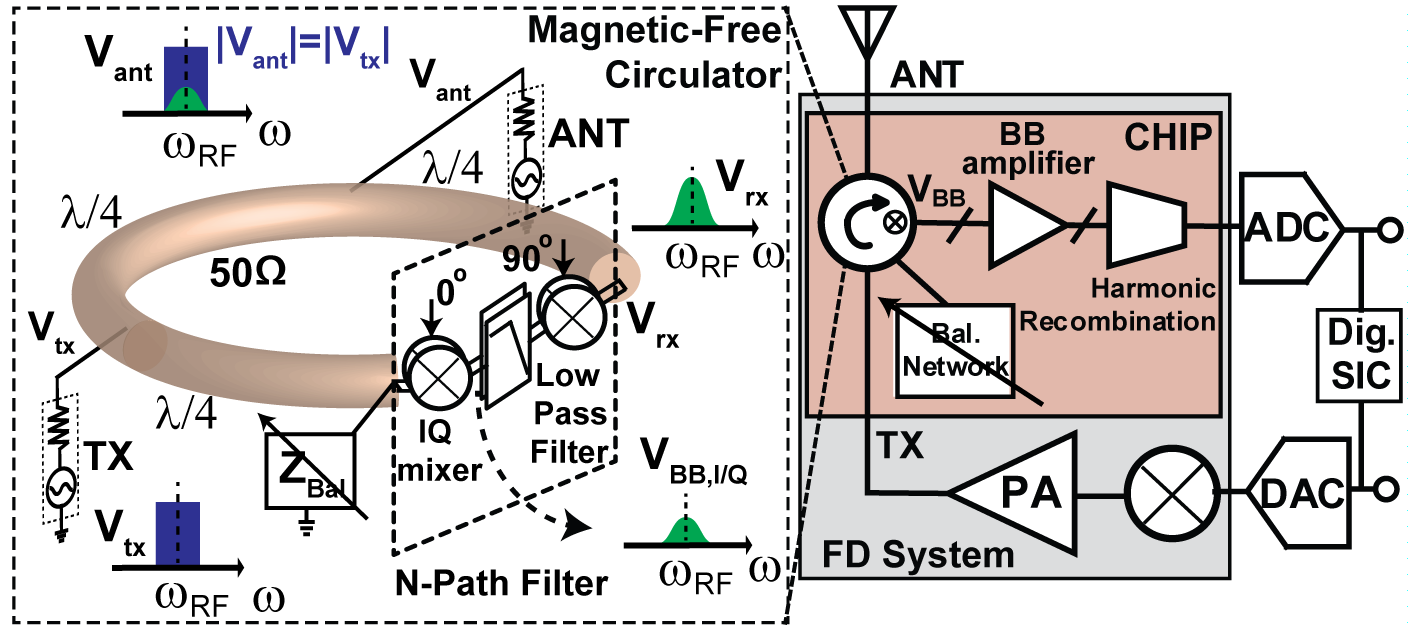}
	\caption{FD N-path-filter-based circulator-receiver conceptual architecture and block diagram.}
	\label{fig:Concept}
	\vspace{-18pt}
\end{figure}

The second challenge is the implementation of fully-integrated single-antenna interfaces which can support simultaneous transmission and reception, while achieving low loss, low noise, high linearity, especially for TX-side excitations, and high isolation between the TX and the RX. Single-antenna FD is important to enable FD capability in hand-held devices \cite{intel_FD_mobile}, as well as extending it to MIMO applications, such as FD massive MIMO base-stations to reduce the overhead of channel state information (CSI) estimation \cite{Ashu_FD_MIMO_CSI}. Previously reported single-antenna FD interfaces either use reciprocal electrical-balance duplexers\cite{EB_Duplexer_Entesari_TMTT2016,imec_TMTT_2016,EB_Duplexer_Larson_TMTT2016}, magnetic circulators \cite{Skyworks_Circulator} or non-reciprocal active quasi-circulators \cite{CornellFullDuplex_JSSC15,ActiveCirculator_RFIC2016}. Reciprocal antenna interfaces suffer from a fundamental minimum of 3dB loss in the TX-ANT and ANT-RX paths. In electrical balance duplexers, the losses can be made asymmetric to favor either TX-ANT or RX NF performance \cite{imec_EBD_unsymmetric}. Ferrite circulators require the use of magnetic materials which renders them incompatible with CMOS fabrication, and therefore bulky and expensive. Active quasi-circulators are limited by the noise and linearity performance of the active devices. 

Reciprocity can be broken using temporal modulation, a fact recognized decades ago in the microwave community \cite{kamal1960parametric, brenner1967unilateral}. More recently, there has been work on non-magnetic passive non-reciprocal circulators based on spatio-temporal modulation of material properties such as permittivity \cite{Ahmed_TMTT,AustinCirculator_TMTT16,NonreciprocalComponents_UCLA_TMTT14} and conductivity \cite{NRK_NatComm16,Tolga_ISSCC2017,TD_NatComm17,UCLA_switch_tline}. These approaches are promising since they can theoretically achieve low loss, low noise, and high isolation, and can be configured to maximize linearity for TX-side excitations as we have shown before in \cite{JZNR_FDradio_ISSCC16,NRK_NatComm16, NRK_JSSC2017}. However, the practical demonstrations of these exciting but nascent approaches currently remain limited in performance, particularly in noise and TX power handling.

In \cite{NRK_NatComm16,JZNR_FDradio_ISSCC16,NRK_JSSC2017}, non-reciprocity is achieved through phase-shifted commutation across a capacitor bank, i.e., within an N-path filter. In \cite{NRK_ISSCC2017}, we introduced a new FD receiver architecture, namely an N-path-filter-based circ.-RX, which re-purposes the commutation-based non-reciprocal circulator to also perform down-conversion and provides direct access to baseband signals at its output, while maintaining noise performance that is comparable to mixer-first receivers \cite{Molnar_JSSC2010,CornellFullDuplex_JSSC15}. The resulting circ.-RX architecture has lower power consumption and NF since the antenna interface and mixing functionalities are achieved within the same block, eliminating the additional LNA/LNTA and mixer and their sources of noise. It also has the additional benefits of allowing the co-design of the antenna interface and the RX, and the embedding of TX-RX isolation-enhancing techniques. The enhancement of TX-RX isolation through the embedding of a reconfigurable balance network increases TX power handling, as the TX swing across the N-path switches is reduced, and more isolation is achieved before the first active baseband amplifier.

This paper expands on \cite{NRK_ISSCC2017} by providing additional information on system requirements and circ.-RX evolution in Section II, analyses to quantify the noise performance, effect of the embedded balance network, and the trade-off between linearity to TX excitations, isolation/balancing range, and TX-ANT/ANT-RX insertion loss in Section III, implementation details of the 65~nm CMOS $0.6-0.9$~GHz circ.-RX in Section IV, and a description of the measurement results and an FD demonstration in Sections V and VI. Finally, Section VII concludes the paper.

\section{FD System Requirements and Circulator-Receiver Evolution} 
\begin{figure}[!t]
	\centering
	\includegraphics[keepaspectratio,width=0.8\linewidth]{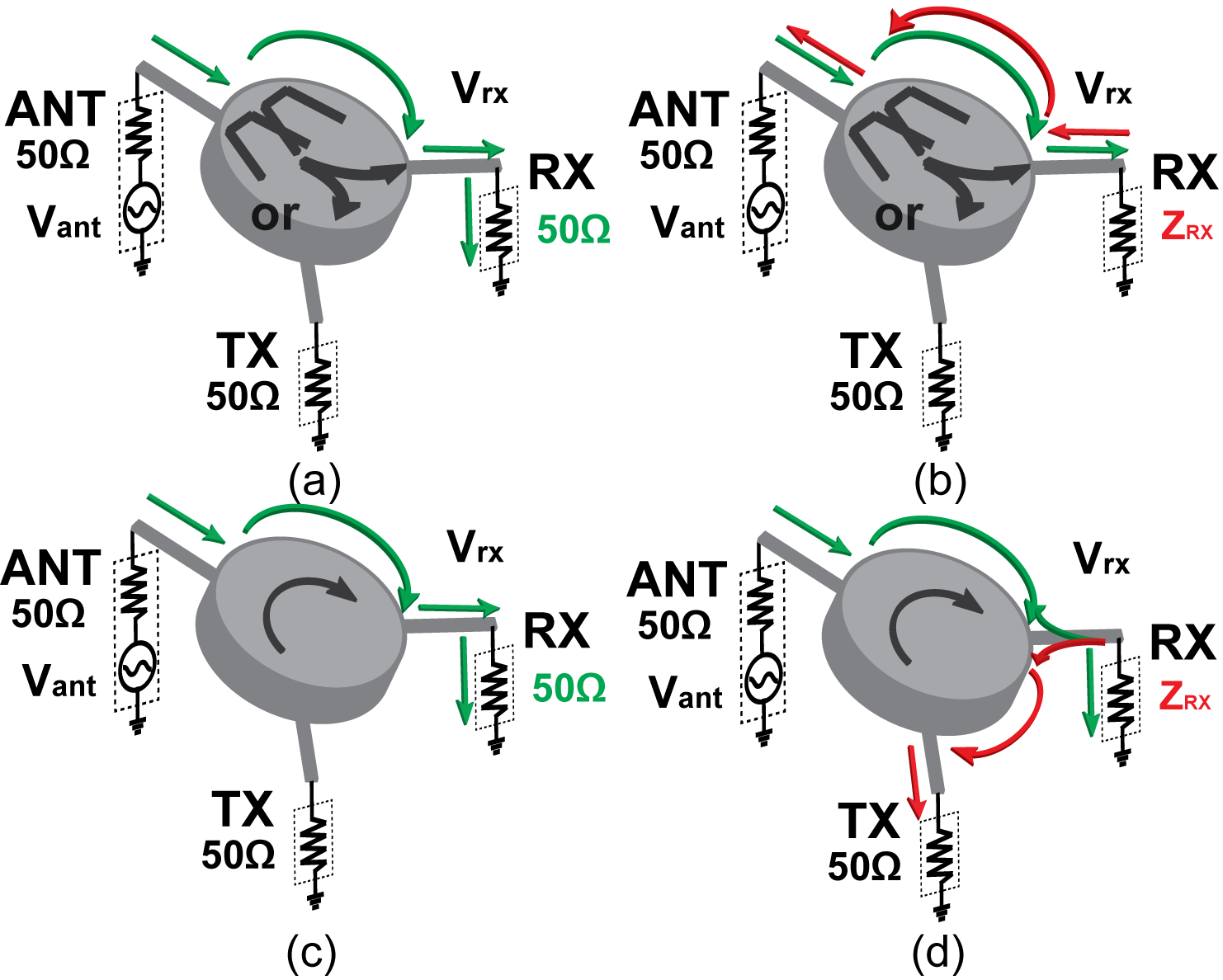}
	\caption{A reciprocal antenna interface: (a) matched at the RX port; (b) unmatched at the RX port where RX reflection returns to the ANT port. A non-reciprocal antenna interface: (a) matched at the RX port; (b) unmatched at the RX port but preserving ANT matching.}
	\label{fig:ANT_matching}
\end{figure}

\begin{figure}[!t]
	\centering
	\includegraphics[keepaspectratio,width=0.75\linewidth]{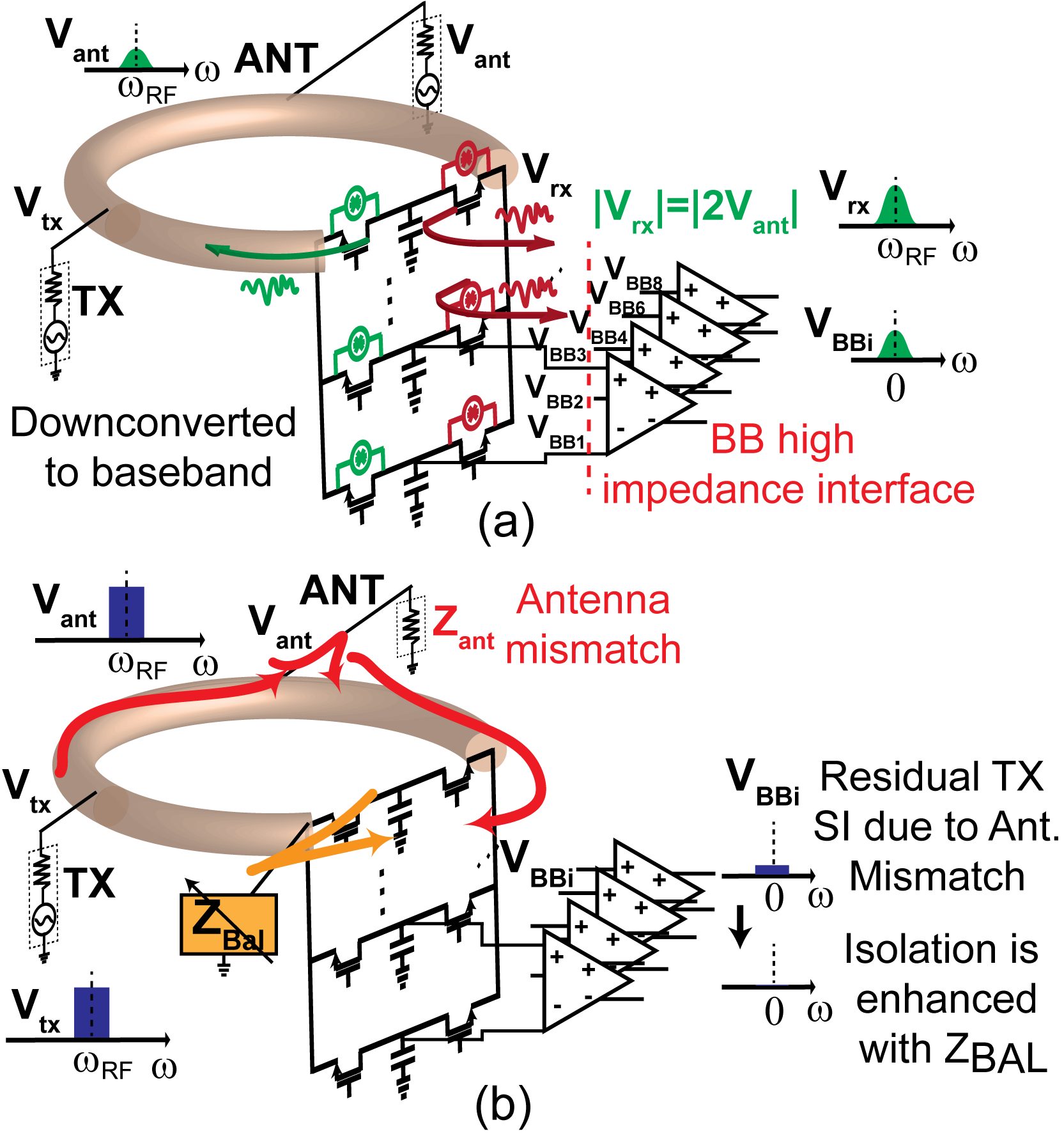}
	\caption{Circulator-receiver features: (a) noise circulation and (b) isolation-enhancement through a balance network that tracks antenna variations.}
	\label{fig:circ_RX_figure}
\end{figure}

\begin{figure}[!h]
	\centering
	\includegraphics[keepaspectratio,width=1.05\linewidth]{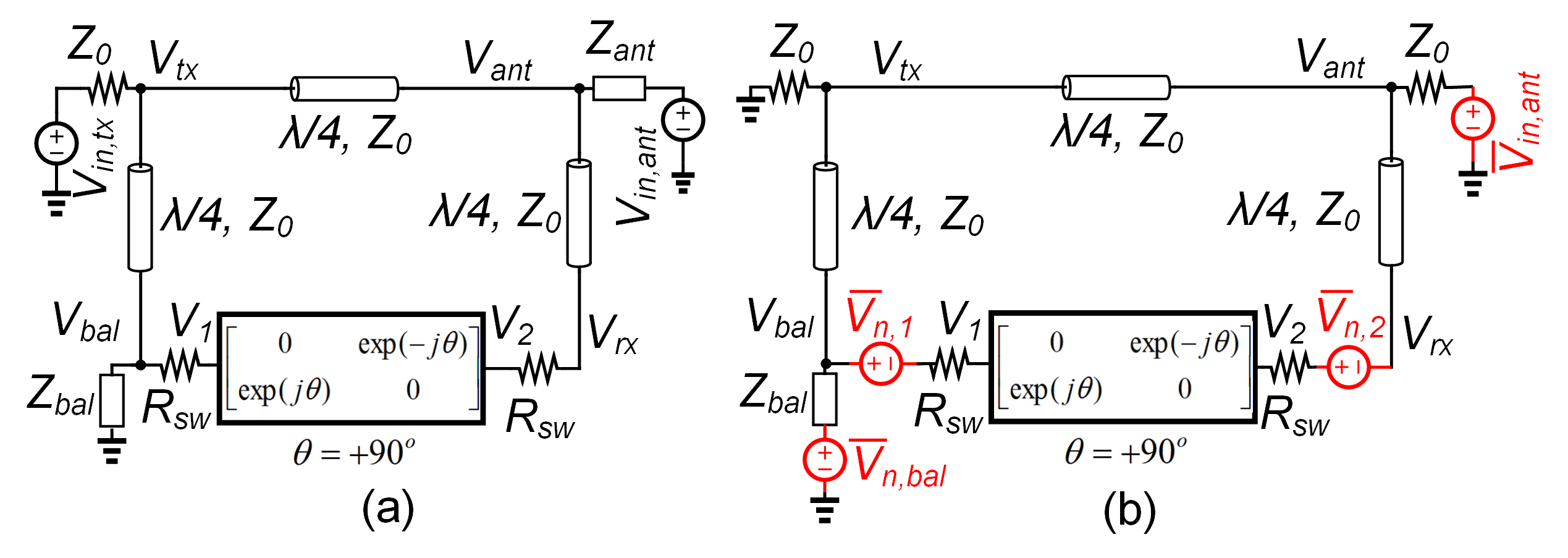}
	\caption{Simplified model to analyze the effect of balance impedance on the performance of the circulator using the analytical S-parameters of the N-path filter (for $N\rightarrow \infty$) and series switch resistances - (a) for TX and ANT excitations, (b) and for noise.}
	\label{fig:Zbal_analysis}
\end{figure}

As mentioned earlier, we have shown that LPTV circuits such as N-path filters enable the achievement of passive non-reciprocity without the use of magnetic materials, and the integration of circulators in CMOS \cite{NRK_NatComm16, NRK_JSSC2017}. This opens new opportunities for the design of fully-integrated FD transceivers where the transceiver is co-designed with the antenna interface. In this section, we define the FD link requirements and discuss some of the interesting features of an FD system where the antenna interface and FD receiver are co-designed and co-optimized as a whole.

\vspace{-6pt}
\subsection{FD Link Requirements}

Consider an FD system with a TX average output power of $+8~$dBm, signal BW of 20~MHz and NF budget of 8dB. The equivalent half-duplex noise floor is $-93~$dBm, requiring $>$100~dB SIC. Our current system, shown in Fig.~\ref{fig:Concept}, is capable of providing about $+80~$dB of SIC across antenna and digital domains, as is discussed later in Section V. Additionally, we have previously reported in \cite{NRK_JSSC2017} the feasibility of BB SIC which can provide an additional $20~$dB cancellation.

Assuming a required SNR of 15~dB, 2.5~dBi TX and RX dipole antenna gain, implementation losses of 5~dB, 10~dB margin for signal fading, and 5~dB (3$\times$) sensitivity degradation due to the residual SI and its IM3, the link budget of $8~\textrm{dBm}+5~\textrm{dBi}-(-93~\textrm{dBm}+5~\textrm{dB}+15~\textrm{dB})-5~\textrm{dB}-10~\textrm{dB}=71~\textrm{dB}$ translates to a transmission distance of 100~meters at a frequency of 750~MHz, which begins to approach the requirements for small-cell FD communication. 

Detailed discussions of linearity requirements for such FD transceivers have been presented previously in \cite{NRK_JSSC2017}. A higher circulator isolation relaxes both the ADC dynamic range and the SI-canceling RX effective IIP3. Additionally, in our circulator architecture, increasing the isolation from TX to RX translates to lower voltage swings on the switches of the non-reciprocal element, which enhances the circulator linearity and transmitter power handling as well.

\vspace{-12pt}
\subsection{RX Matching in Non-Reciprocal Antenna Interfaces}

Integrated receivers are typically designed to have a $50~\Omega$ input impedance to provide matching for the conventional reciprocal antenna interfaces that precede them, such as SAW duplexers and filters. Matching is necessary to obtain best filtering performance from the duplexer or filter. Additionally, any mismatch at the receiver port causes a reflection which travels back to the antenna due to reciprocity and causes undesired ANT re-radiation (Figs.~\ref{fig:ANT_matching}(a)-(b)). The need to simultaneously achieve input matching to $50~\Omega$ and noise matching for low noise performance, particularly over wide bandwidths, is the fundamental challenge of LNA/LNTA design.

Our FD receiver in \cite{NRK_JSSC2017} followed a similar conventional design methodology, where the integrated circulator and the subsequent receiver were both designed to provide a $50~\Omega$ impedance at the RX port. However, in the case of non-reciprocal antenna interfaces such as circulators, the receiver reflection circulates away from the ANT and is absorbed at the TX port (Fig.~\ref{fig:ANT_matching}(c)-(d)). Hence, the $Z_{rx}$ value shown in Fig.~\ref{fig:ANT_matching}(d) does not need to be $50~\Omega$. It can be shown that increasing $Z_{rx}$ increases the voltage gain from ANT to the RX, and for the case of a high impedance RX interface ($Z_{rx}\rightarrow \infty$), the voltage gain is maximized to $6~$dB for an ideal circulator.

\begin{equation}
|V_{rx}|=|\frac{2V_{ant}}{1+\frac{50}{Z_{rx}}}|, \quad \textrm{for} \quad Z_{rx}\rightarrow \infty: |V_{rx}|=|2V_{ant}|.
\label{eq:Zrx_gain}
\end{equation}

It is worth pointing out that a non-reciprocal antenna interface can provide additional benefits such as lower LO and harmonic re-radiation at the ANT port, since these signals circulate away from the antenna as well. 

\subsection{Embedded Down-Conversion and Noise Circulation}

A conceptual architecture and block diagram of the circulator-receiver are presented in Fig.~\ref{fig:Concept}. The circulator consists of a non-reciprocal LPTV gyrator built using an N-path filter \cite{NRJZ_TCAS2}, combined with three transmission line sections of an overall length of $3\lambda /4$ at the operation frequency \cite{NRK_JSSC2017,NRK_NatComm16}. As with all N-path filters, the non-reciprocal element can also be viewed as a mixer-low pass filter (LPF)-mixer structure with phase-shifted clocks. In such a structure, it can be seen intuitively that any RF signal appearing on $Z_{rx}$ is down-converted and low-pass-filtered by the BB capacitors. In other words, the circulator structure inherently includes a mixer-first receiver, and the BB signals captured on the BB capacitors can be used for further BB processing. Importantly, isolation continues to be seen between the TX port and the N-path filter BB nodes. 

\begin{figure}[!t]
	\centering
	\includegraphics[keepaspectratio,width=1\linewidth]{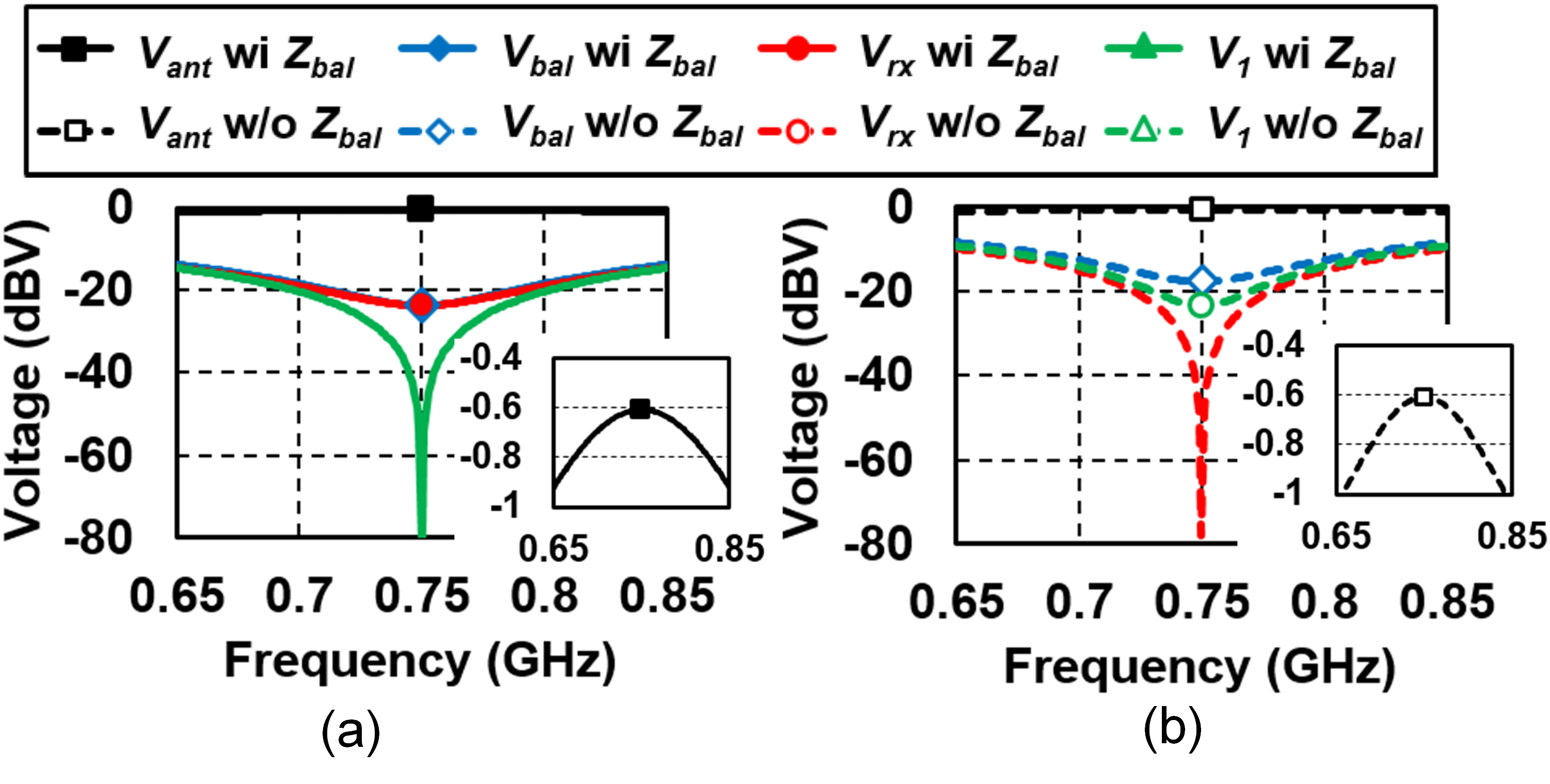}
	\caption{Analysis and simulation results of the ideal circ.-RX shown in Fig.~\ref{fig:Zbal_analysis}(a), (a) without and (b) with the balance network across frequency for $V_{in,tx}=2~V$. The analysis is shown with markers while the lines represent simulations across frequency. $Z_{ant}=50~\Omega$, $R_{sw}=3.5~\Omega$ and $Z_{ant}=50~\Omega$.}
	\label{fig:Zbal_sims}
\end{figure}

\begin{figure}[!h]
	\centering
	\includegraphics[keepaspectratio,width=0.9\linewidth]{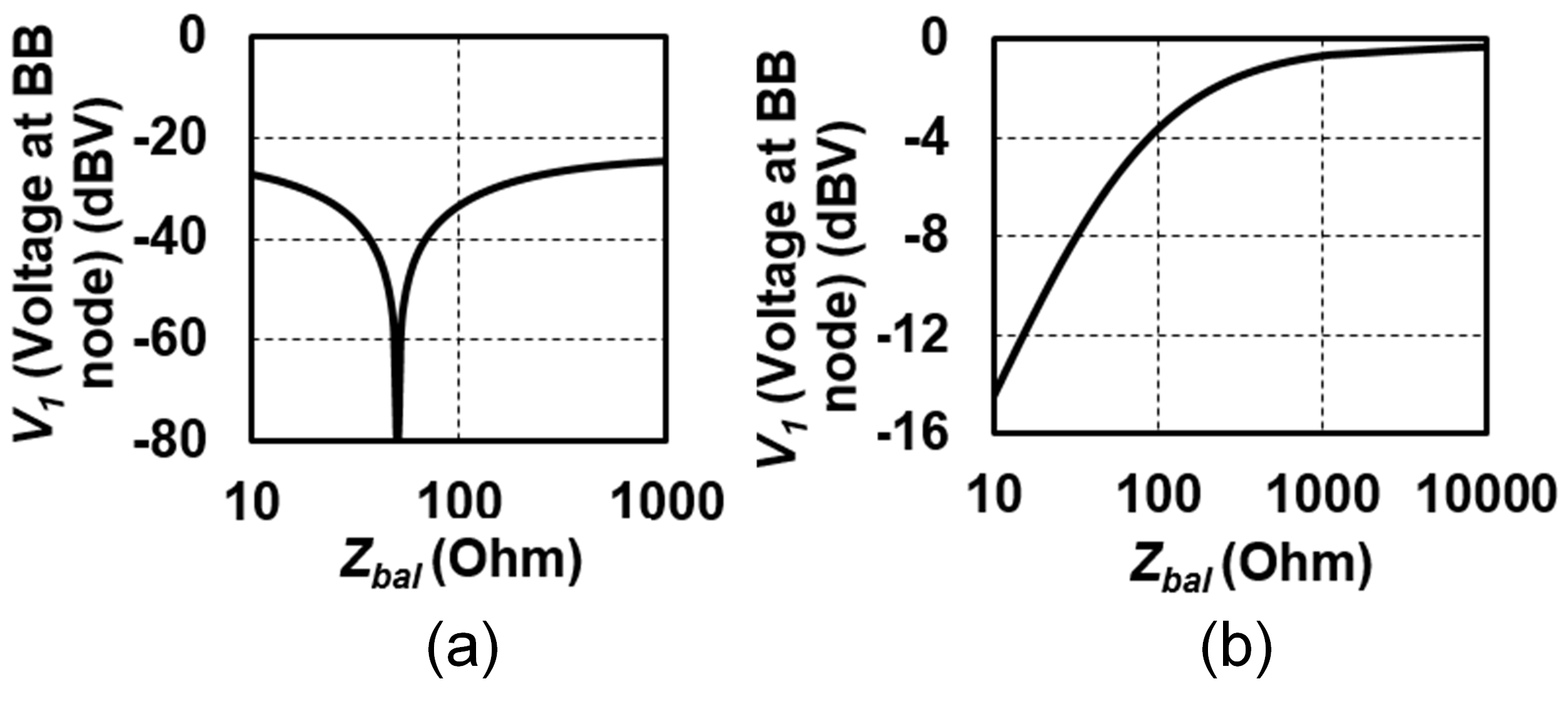}
	\caption{Analysis of the effect of balance network on the ideal circ.-RX: gyrator voltage $V_1$, representing the BB nodes, as a function of the balance resistance (a) for a $V_{in,tx}=2~V$ TX excitation in Fig.~\ref{fig:Zbal_analysis}(a) ($V_{in,ant}=0~V$), and (b) for a $V_{in,ant}=2~V$ ANT excitation in Fig.~\ref{fig:Zbal_analysis}(a) ($V_{in,tx}=0~V$). $Z_{ant}=50~\Omega$ and $R_{sw}=3.5~\Omega$.}
	\label{fig:Zbal_sims2}
\end{figure}

Recall that in a passive non-return-to-zero (NRZ) mixer, we have \cite{RF_razavi}:
\begin{equation}
\frac{I_{BB,1-path}}{I_{RF,in}}=\frac{2}{\pi}\frac{sin(\pi D)}{2D}.
\label{eq:passive_mixer}
\end{equation}

\noindent where $I_{BB1-path}$ and $I_{RF,in}$ are the currents flowing into one of the baseband paths and from the RF source, respectively. As an example, for an 8-path NRZ mixer with $D=12.5\%$, $\frac{I_{BB1-path}}{I_{RF,in}}=-0.22~$dB, compared to $20log(\frac{2}{\pi})=-3.92$~dB in a 2-path mixer. In other words, similar to the case of the NRZ mixer, by having a higher number of paths in the N-path filter, $N$, with lower duty cycle, $D=\frac{1}{N}$, the RF-to-IF loss is improved since the fundamental component of the Fourier series for each LO waveform is stronger. Subsequent recombination of different phases to create the I/Q signal can provide an effective RF-to-IF voltage gain. For example, combining the 8 phases for 3rd and 5th harmonic rejection with $\pm1,\pm\frac{1}{\sqrt2},0,\mp\frac{1}{\sqrt2}$ weights for 0$^\circ$/180$^\circ$, 45$^\circ$/225$^\circ$, 90$^\circ$/270$^\circ$ and 135$^\circ$/315$^\circ$ phases results in an additional 12~dB recombination voltage gain. 

\begin{figure}[!t]
	\centering
	\includegraphics[keepaspectratio,width=0.65\linewidth]{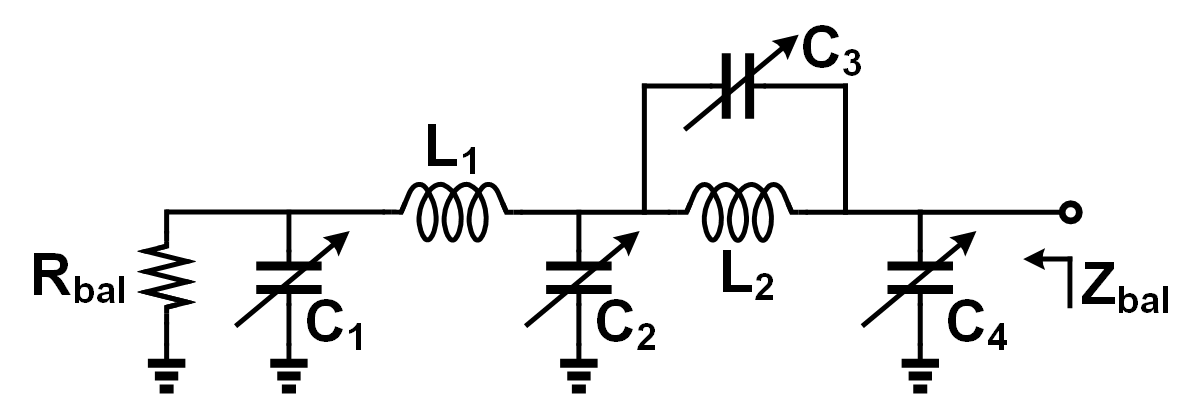}
	\caption{Balance network structure used in simulations \cite{imec_TMTT_2016}.}
	\label{fig:Zbal_EBD}
\end{figure}

\begin{figure}[!h]
	\centering
	\includegraphics[keepaspectratio,width=1\linewidth]{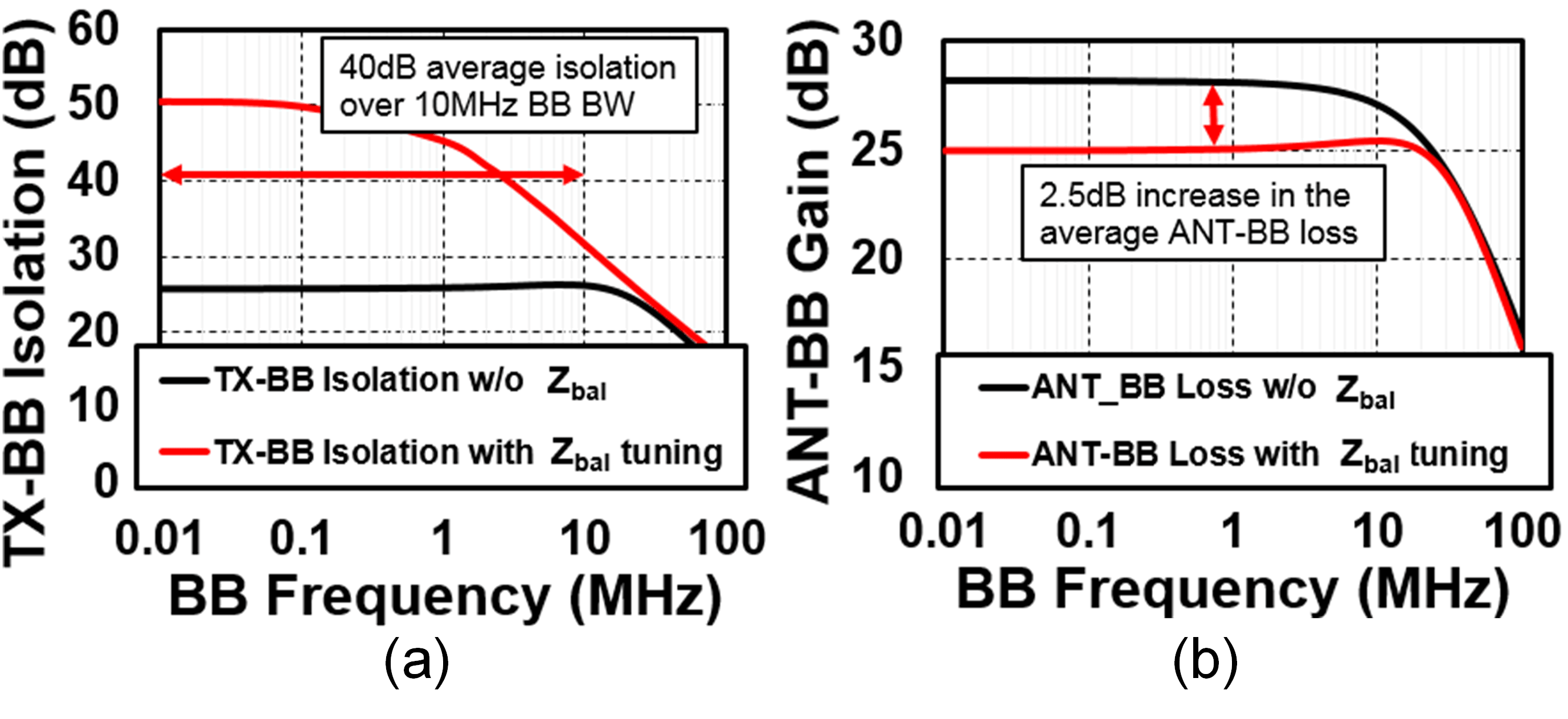}
	\caption{Simulated (a) TX-BB isolation and (b) ANT-BB gain of the circulator receiver. Simulations are run on the extracted post-layout schematics of the circulator-receiver, including wirebond inductance and package parasitics, with the LC ladder-based balance network (inductor Q=20) model shown in Fig.~\ref{fig:Zbal_EBD}.}
	\label{fig:Zbal_sims_Cadence}
\end{figure}

An additional interesting behavior in the circ.-RX is noise circulation. As we will show in Section III, the noise of the RX-side switches contribute mainly to the RX NF while the noise of the TX-side switches circulates away. Hence, the NF of the circ.-RX is theoretically as low as that of traditional mixer-first RXs despite the additional set of switches (Fig.~\ref{fig:circ_RX_figure}(a)). 

\subsection{Enhancing Circulator Isolation}

The TX-to-RX isolation in any three-port shared-antenna interface is limited to the quality of matching at the antenna port. In a practical system, antenna matching depends heavily on environmental reflections. As a result, antenna tuners are necessary to maintain TX-to-RX isolation across ANT variations. If placed at the antenna port, the tuner should be as linear as the antenna interface itself for TX signals. In addition, parasitics within the circulator itself can limit the isolation achieved. Inspired by the concept of the balance network in electrical-balance duplexers \cite{imec_TMTT_2016}, we have found that incorporating a tunable impedance on the TX side of the N-path filter, as shown in Fig.~\ref{fig:circ_RX_figure}(b), can enhance TX-RX isolation. In essence, the tunable impedance creates a reflection that cancels out the reflected TX signal leaking to the BB nodes. It is also notable that in this work, the balance network is placed at a low-voltage-swing node with respect to the TX, hence maintaining the linearity benefits of the circulator. However, there exists a trade-off between linearity, isolation, and loss, as will be discussed in the following section. In this work, we have prioritized linearity and TX power handling, and therefore the balance network is mainly effective in combating the effect of circuit parasitics, rather than handling large ANT variations.

\begin{figure}[!t]
	\centering
	\includegraphics[keepaspectratio,width=1\linewidth]{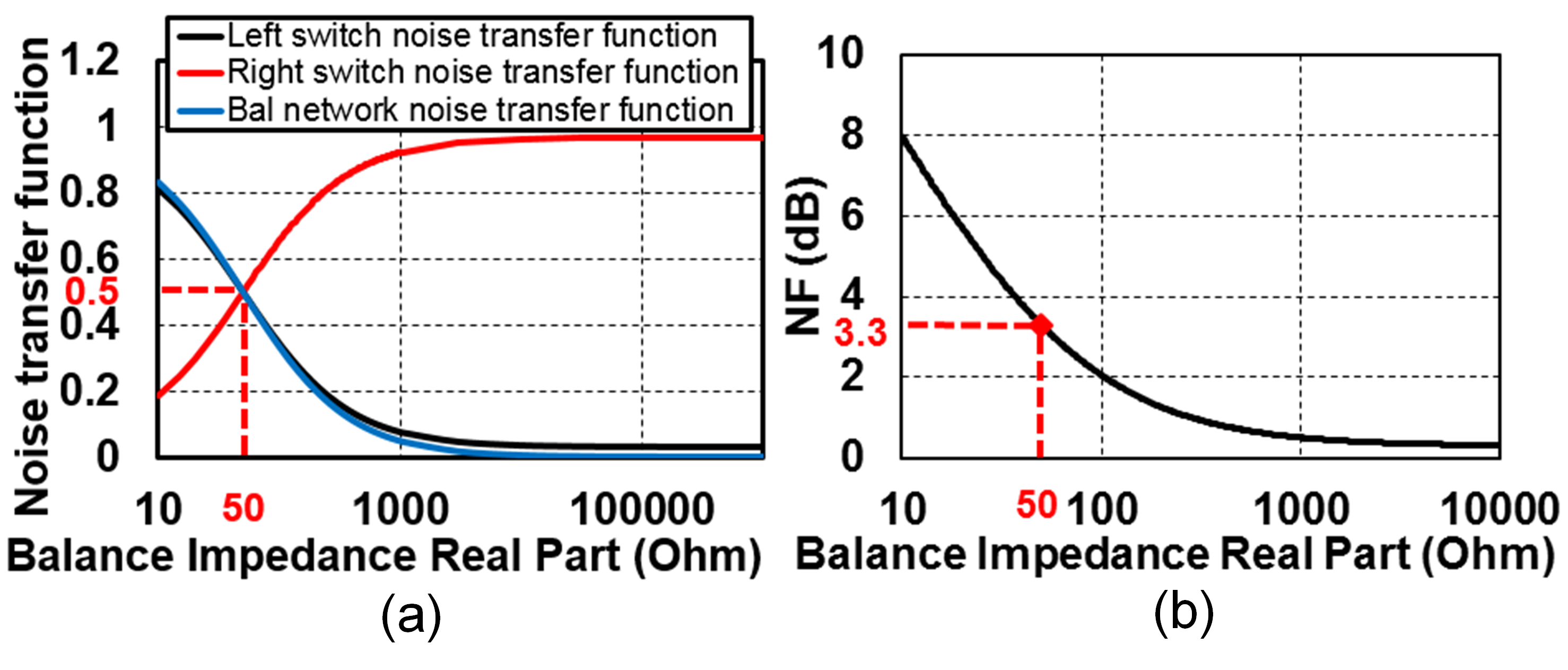}
	\caption{(a) Noise transfer functions based on (\ref{eq:Noise_circulation_withZbal}) and (b) theoretical NF of the circulator-receiver in the presence of the balance network.}
	\label{fig:NTF}
\end{figure}

\begin{figure}[!h]
	\centering
	\includegraphics[keepaspectratio,width=0.5\linewidth]{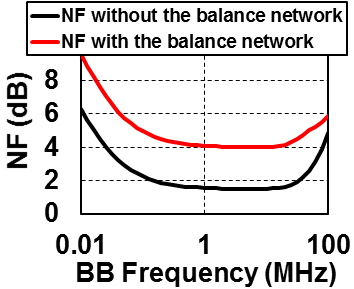}
	\caption{Simulated NF of the circulator-receiver with and without the balance network optimized for 40~dB average isolation. The settings are similar to those of the simulation results shown in  Fig.~\ref{fig:Zbal_sims_Cadence}.}
	\label{fig:NF_circRX}
\end{figure}
\section{Analysis of Integrated Commutation-Based Circulator-Receiver Performance Metrics}
\subsection{Effect of the Balance Network}

In \cite{NRK_JSSC2017}, we demonstrated a simplified method of analyzing the N-path-filter-based circulator, by using the fundamental-to-fundamental S parameters of the N-path filter as a simplified model (ignoring harmonic conversion effects) along with conventional microwave circuit analysis techniques for the overall structure. Using the same method, a simplified model for analyzing the circulator-receiver in the presence of the balance network is shown in Fig.~\ref{fig:Zbal_analysis}(a). The N-path filter is modeled using an ideal gyrator (assuming a large number of paths $N$) and the effect of switch resistance is captured through two series resistances with the gyrator element.

For an excitation at the TX, $V_{in,tx}$ in Fig.~\ref{fig:Zbal_analysis}, the voltages at various nodes within the circulator can be found as:

\begin{equation}
V_{ant}=-jV_{in,tx}\frac{Z_0}{R_{sw}}\frac{1}{2+\frac{Z_0}{R_{sw}}\left(1+\frac{Z_0}{Z_{ant}}\right)}
\label{eq:TX_excitation_1}
\end{equation}

\begin{equation}
V_{bal}=-jV_{in,tx}\frac{2+\frac{Z_0}{R_{sw}}\left(\frac{Z_0}{Z_{ant}}-1\right)}{\left(1+\frac{Z_0}{Z_{bal}}\right)\left(2+\frac{Z_0}{R_{sw}}\left(1+\frac{Z_0}{Z_{ant}}\right)\right)}
\label{eq:TX_excitation_2}
\end{equation}

\begin{equation}
V_{1}=-jV_{2}=-jV_{in,tx}\frac{1-\frac{Z_0}{Z_{bal}}+\frac{Z_0}{R_{sw}}\left(\frac{Z_0}{Z_{ant}}-1\right)}{\left(1+\frac{Z_0}{Z_{bal}}\right)\left(2+\frac{Z_0}{R_{sw}}\left(1+\frac{Z_0}{Z_{ant}}\right)\right)}
\label{eq:TX_excitation_3}
\end{equation}

\begin{figure}[!t]
	\centering
	\includegraphics[keepaspectratio,width=1\linewidth]{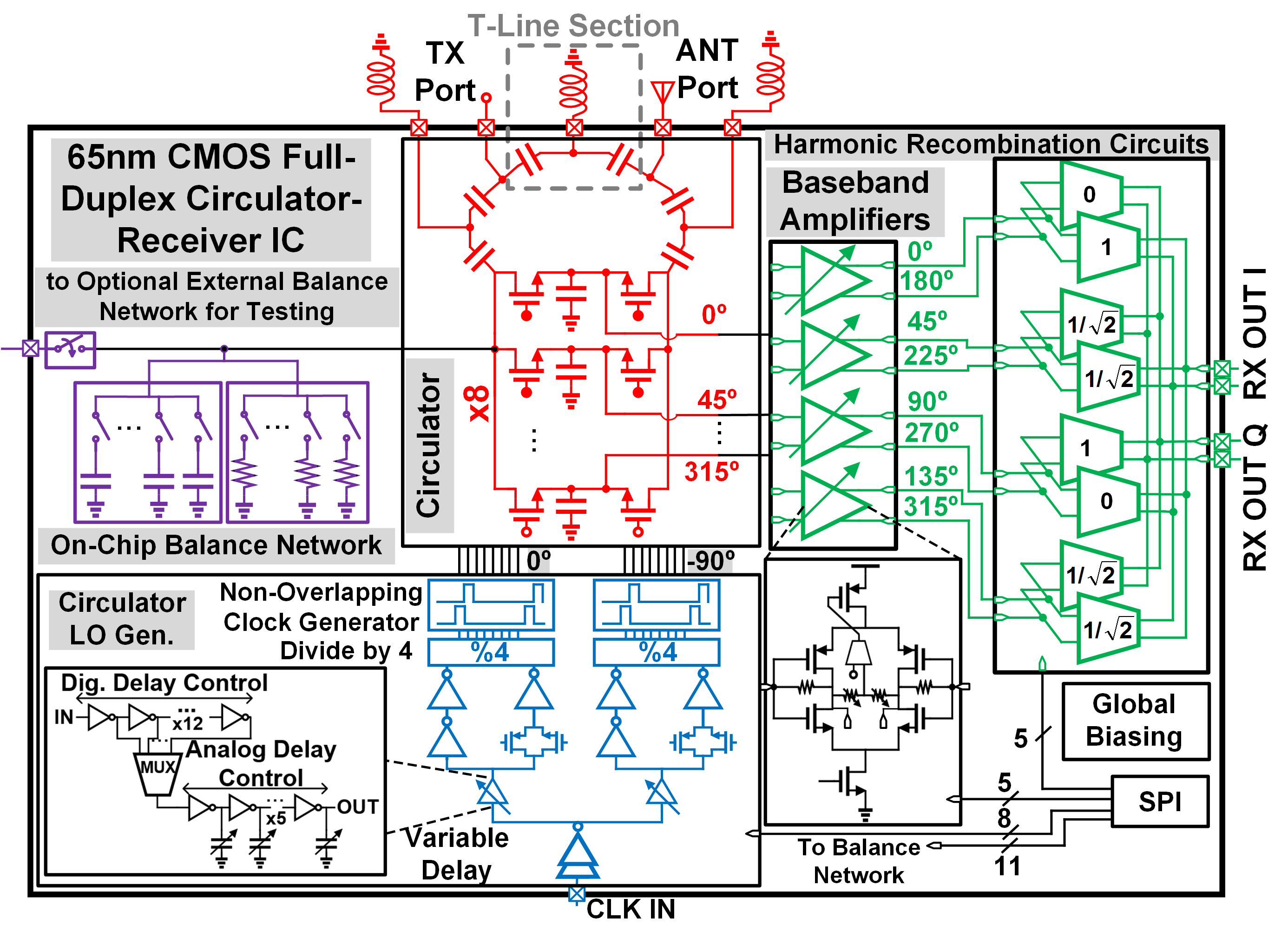}
	\caption{Block diagram and schematic of the implemented 65~nm CMOS FD circulator-RX with integrated circulator and balancing impedance.}
	\label{fig:chip_block_diagram}
\end{figure}

\begin{figure}[!h]
	\centering
	\includegraphics[keepaspectratio,width=0.65\linewidth]{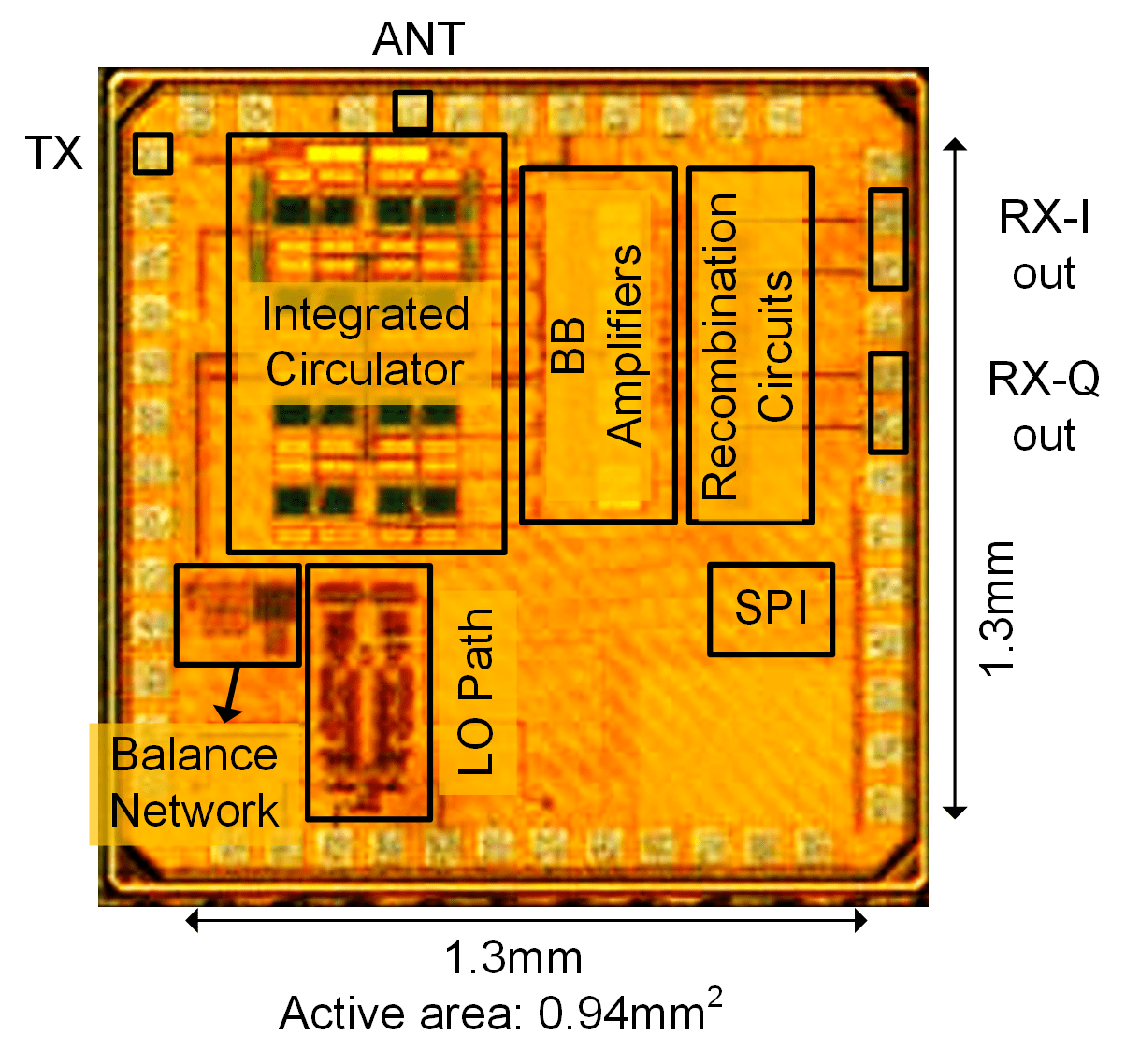}
	\caption{Chip microphotograph of the 65~nm CMOS full-duplex circulator-receiver.}
	\label{fig:Chip_Photo}
\end{figure}

\begin{equation}
V_{rx}=V_{in,tx}\frac{-\frac{2Z_0}{Z_{bal}}+\frac{Z_0}{R_{sw}}\left(\frac{Z_0}{Z_{ant}}-1\right)}{\left(1+\frac{Z_0}{Z_{bal}}\right)\left(2+\frac{Z_0}{R_{sw}}\left(1+\frac{Z_0}{Z_{ant}}\right)\right)}
\label{eq:TX_excitation_4}
\end{equation}

\noindent where $V_{ant}$ is the voltage at the antenna port, $V_{rx}$ and $V_{bal}$ are the voltages at the right and left sides of the simplified non-reciprocal N-path filter in Fig.~\ref{fig:Zbal_analysis}(a), and $V_1$ and $V_2$ are the voltages at the ideal gyrator ports (essentially the BB nodes, but without the frequency translation effect). This analysis in its current form is only valid at the center frequency of operation. However, an approach similar to \cite{NRK_JSSC2017} can be used to model the variation of the transmission lines' response across frequency, as well as that of the N-path filter. Due to the complexity of the equations, here we opted to limit the analysis to the center frequency and use Cadence PSS+PAC simulations to verify our results and show the performance across frequency. 

\begin{figure}[!t]
	\centering
	\includegraphics[keepaspectratio,width=0.7\linewidth]{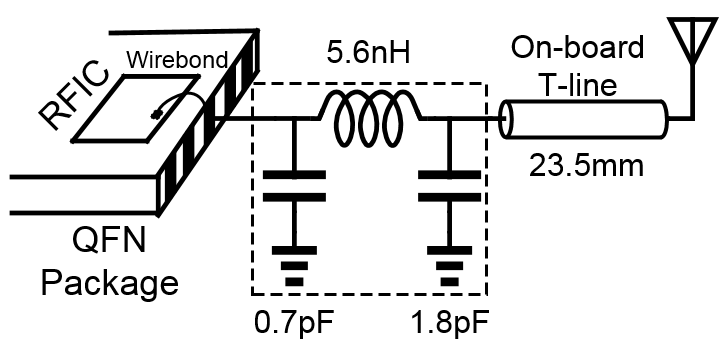}
	\caption{The SMD CLC-based ANT matching network used on the PCB to compensate for parasitics and achieve a reasonable nominal TX-to-RX isolation.}
	\label{fig:matching_network}
\end{figure}

\begin{figure*}[!t]
	\centering
	\includegraphics[keepaspectratio,width=0.85\linewidth]{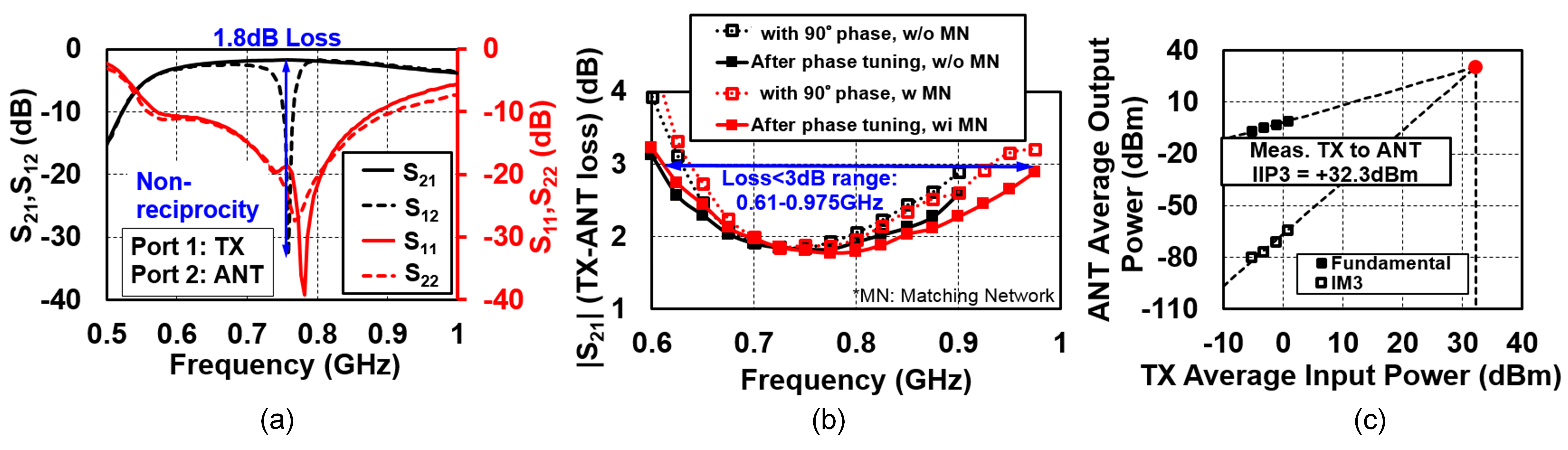}
	\caption{(a) Circulator TX-to-ANT S-parameter measurements for 750~MHz clock frequency. (b) Measured TX-to-ANT loss at the center frequency as the clock frequency is tuned. Phase tuning is used at each frequency to ensure minimum loss. The black/red curves are without/with the SMD-based ANT matching network. (c) Measured IB TX-to-ANT IIP3.}
	\label{fig:circ_SP_IIP3}
\end{figure*}

From Eq. (\ref{eq:TX_excitation_3}), by setting $V_{1}$ to 0, a formula for the desired $Z_{bal}$ can be derived which nulls the TX leakage at the gyrator as follows:

\begin{equation}
Z_{bal}=\frac{Z_0R_{sw}Z_{ant}}{R_{sw}Z_{ant}+Z_0(Z_0-Z_{ant})}
\label{eq:TX_excitation_5}
\end{equation}

\noindent Various important points can be deduced from the equations above. Firstly, $Z_{bal}$ becomes 0 when $R_{sw}$ is 0. This is representative of the fact that when $R_{sw}$ is 0, the gyrator is ideal, which means that if balancing has been accomplished and $V_{1}$ is 0, then $V_{bal}$ is 0 as well. This would render a shunt impedance at $V_{bal}$ ineffectual. \emph{Therefore, while placing the balancing impedance at $V_{bal}$ leads to linearity benefits due to the low swing at that node for TX excitations, it limits the utility of the balancing impedance to the compensation of parasitics within the circulator and minor ANT impedance variations}. For $R_{sw}\neq 0$, for $Z_{ant}=50~\Omega$, the required $Z_{bal}$ that results in perfect isolation is also $50~\Omega$. Fig.~\ref{fig:Zbal_sims} shows the results of our analysis at the center frequency as well as simulations across frequency using Cadence for the ideal circuit analyzed above without the balance network and with the addition of the balance network for an excitation at the TX port. As can be seen, without the balance network, in the presence of finite $R_{sw}$, perfect isolation is seen at $V_{rx}$ but there is finite voltage swing at $V_{1}$ and $V_{BAL}$. However, by tuning the balance network according to Eq.~(\ref{eq:TX_excitation_5}) (i.e. to $50~\Omega$), the TX leakage is cancelled at the gyrator node $V_{1}$. Secondly, the TX-to-ANT loss in this case is largely not affected by the balance impedance $Z_{bal}$, and only depends on the $R_{sw}$ and $Z_{ant}$. \emph{This is another expected benefit of placing the balancing impedance at a node which has low swing for TX signals.} Thirdly, as an example after balancing, $V_{bal}$ is -23~dB below $V_{tx}$ for $R_{sw}=3.5\Omega$ and $Z_{ant}=50\Omega$, indicating the extent to which the power handling requirements on the balance network are relaxed. 

Similarly, for an excitation at the ANT port, $V_{in,ant}$, the various voltages are:
\begin{equation}
V_{1}=-jV_{2}=-2V_{in,ant}\frac{\frac{Z_0}{Z_{ant}}\left(\frac{Z_0}{R_{sw}}+\frac{1}{2}\left(1+\frac{Z_0}{Z_{bal}}\right)\right)}{\left(1+\frac{Z_0}{Z_{bal}}\right)\left(2+\frac{Z_0}{R_{sw}}\left(1+\frac{Z_0}{Z_{ant}}\right)\right)}
\label{eq:ANT_excitation_1}
\end{equation}

\begin{equation}
V_{bal}=-2V_{in,ant}\frac{\frac{Z_0}{Z_{ant}}\frac{Z_0}{R_{sw}}}{\left(1+\frac{Z_0}{Z_{bal}}\right)\left(2+\frac{Z_0}{R_{sw}}\left(1+\frac{Z_0}{Z_{ant}}\right)\right)}
\label{eq:ANT_excitation_2}
\end{equation}

\begin{equation}
V_{rx}=-2V_{in,ant}\frac{\frac{Z_0}{Z_{ant}}\left(\frac{Z_0}{R_{sw}}+\left(1+\frac{Z_0}{Z_{bal}}\right)\right)}{\left(1+\frac{Z_0}{Z_{bal}}\right)\left(2+\frac{Z_0}{R_{sw}}\left(1+\frac{Z_0}{Z_{ant}}\right)\right)}
\label{eq:ANT_excitation_3}
\end{equation}

\begin{table*}[!t]
	\begin{equation}
	\overline{V^2_1}=\overline{V^2_2}=\frac{1}{4}\overline{V^2_{n,ant}}\left(1+\frac{\Gamma_{bal}}{1+\frac{R_{sw}}{Z_0}}\right)^2+\frac{1}{4}\overline{V^2_{n,1}}\left(1-\frac{\Gamma_{bal}}{1+\frac{R_{sw}}{Z_0}}\right)^2+\frac{1}{4}\overline{V^2_{n,2}}\left(1+\frac{\Gamma_{bal}}{1+\frac{R_{sw}}{Z_0}}\right)^2+\frac{1}{4}\overline{V^2_{n,bal}}\left(1-\Gamma_{bal}\right)^2
	\label{eq:Noise_circulation_withZbal}
	\end{equation}
\end{table*}

\begin{table*}[!t]
	\vspace{-12pt}
	\begin{equation}
	F=1+\frac{R_{sw}}{Z_0}\left(\frac{1+\frac{R_{sw}}{Z_0}-\Gamma_{bal}}{1+\frac{R_{sw}}{Z_0}+\Gamma_{bal}}\right)^2+\frac{R_{sw}}{Z_0}+\frac{Re(Z_{bal})}{Z_0}\left(\frac{(1+\frac{R_{sw}}{Z_0})(1-\Gamma_{bal})}{1+\frac{R_{sw}}{Z_0}+\Gamma_{bal}}\right)^2.
	\label{eq:Noise_figure}
	\end{equation}
	\hrule
	\vspace{-12pt}
\end{table*}

Fig.~\ref{fig:Zbal_sims2} shows the trade-off between TX-BB isolation and ANT-RX loss by plotting $V_{1}$  using the equations above for both TX and ANT excitations and different values of the balancing resistance ($Z_{ant}=50~\Omega$ and $R_{sw}=3.5~\Omega$). It can be seen that unlike the previous case, the ANT-to-gyrator loss increases with a lowering of the magnitude of $Z_{bal}$. \emph{Therefore, there exists a trade-off between TX-BB isolation and ANT-BB loss using this balancing scheme.}

The aforementioned approximate analysis captures the basics of how the balancing network functions in the circ.-RX, and shows how it can compensate for switch resistance parasitics. To verify this intuitive understanding for other types of parasitics, we have simulated our circ.-RX, along with an LC ladder-based balance network model as used in \cite{imec_TMTT_2016}, also shown in Fig.~\ref{fig:Zbal_EBD}. This impedance network can provide orthogonal impedance tunability just by varying the capacitor values. The inductors are fixed with a Q of 20. The circulator-receiver contains extracted layout parasitics, wirebond inductance and package parasitics. Fig.~\ref{fig:Zbal_sims_Cadence} shows the ANT-to-BB gain and the TX-to-BB isolation with and without the balance network, assuming 50~$\Omega$ antenna impedance. As mentioned earlier, the balancing network is being used to compensate for layout parasitics, wirebond inductance and package parasitics. More than 40dB average isolation can be achieved over 20MHz RF BW with a 2.5~dB degradation in the ANT-RX loss. In our implementation, discussed later in this paper, we have implemented a balance network consisting only of tunable capacitors and resistors, thanks to an SMD CLC-based fixed ANT tuning network incorporated on the printed circuit board (PCB) performing a minor transformation (VSWR=1.2).

\subsection{Noise Circulation}

Here, we derive equations for the noise transfer function of the switch noise sources as well as the balance network. The circuit diagram for noise analysis is shown in Fig.~\ref{fig:Zbal_analysis}(b) where the various noise sources are highlighted in red. The overall noise voltage at the gyrator is shown in Eq. (\ref{eq:Noise_circulation_withZbal}) (assuming equal switch resistance on both sides of the gyrator). Where $\Gamma_{bal}$ is the reflection coefficient of the balance network, $\overline{V_{n,ant}}=4KTZ_0$, $\overline{V_{n,i}}=4KTR_{sw,i}$ ($i= 1,2$) and $\overline{V_{n,bal}}=4KT.Re(Z_{bal})$. This results in an equivalent noise figure as in Eq. (\ref{eq:Noise_figure}).

\begin{figure}[!t]
	\centering
	\includegraphics[keepaspectratio,width=\linewidth]{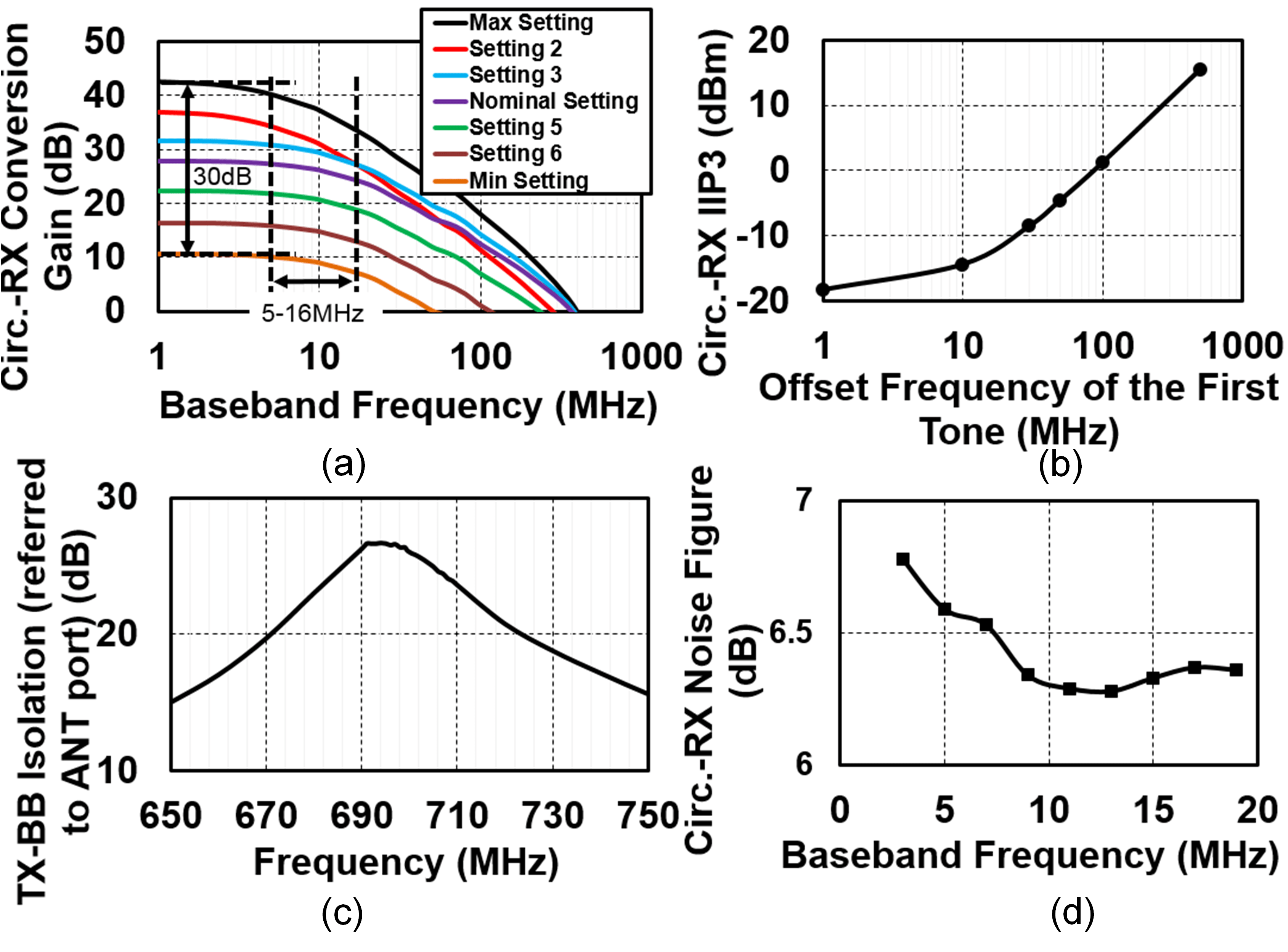}
	\caption{Measured circulator-receiver (a) conversion gain, (b) IIP3, (c) small-signal TX-to-BB isolation for -40~dBm TX power and (d) NF. The balance network is not engaged in these measurements.}
	\label{fig:RX_meas}
\end{figure}

\noindent It should be mentioned that Eq. (\ref{eq:Noise_figure}) only captures the noise of the circulator alone and does not include the noise of the following BB circuitry. Additionally, the noise equations are all derived for $Z_{ant}=Z_0$, but can be modified to include the effect of varying antenna impedance. Since we are not working with large ANT variations in this work, we have used the nominal antenna impedance for simplicity here. 

From these equations, it can be seen that the noise of the left-hand-side switches contributes differently to the noise voltage at the gyrator than the noise from the antenna and the right-hand-side switches. Assuming that $R_{sw}$ is relatively small compared to $Z_0$, it can be seen from (\ref{eq:Noise_circulation_withZbal}) that the noise from the left-hand-side switches vanishes at the gyrator node when there's no balance network. In the presence of the balance network, the noise transfer functions from the switches depend on the network's reflection coefficient. Fig.~\ref{fig:NTF} shows the noise transfer functions and the  NF calculated above as a function of the balance network impedance for $R_{sw}=3.5\Omega$ (no other post-layout parasitics included). It should be mentioned that this NF plot is the worst case scenario, assuming that the balance network is completely resistive. As the balance network impedance becomes smaller, the noise contribution from the left-hand-side switches and the balance network increases. For the optimal isolation ($Z_{bal} = 50\Omega$, as calculated earlier) , the NF degrades by about $3$~dB. This shows the trade-off between the TX-to-BB isolation and RX NF. Fig.~\ref{fig:NF_circRX} shows NF simulations of the circ.-RX for $Z_{ant}=50\Omega$ with and without the balance network compensating for post-layout and package parasitics. The overall NF is degraded by around 2.5~dB to 4dB. Our measured NF is higher due to the additional noise contributions from the BB amplifiers and the clock circuitry. An interesting future expansion of this noise analysis is to analyze the effect of phase noise on the circulator, as has recently been done for N-path filters in \cite{Cornell_PN_TCAS}.

\section{Implementation}
\begin{figure}[!t]
	\centering
	\includegraphics[keepaspectratio,width=1\linewidth]{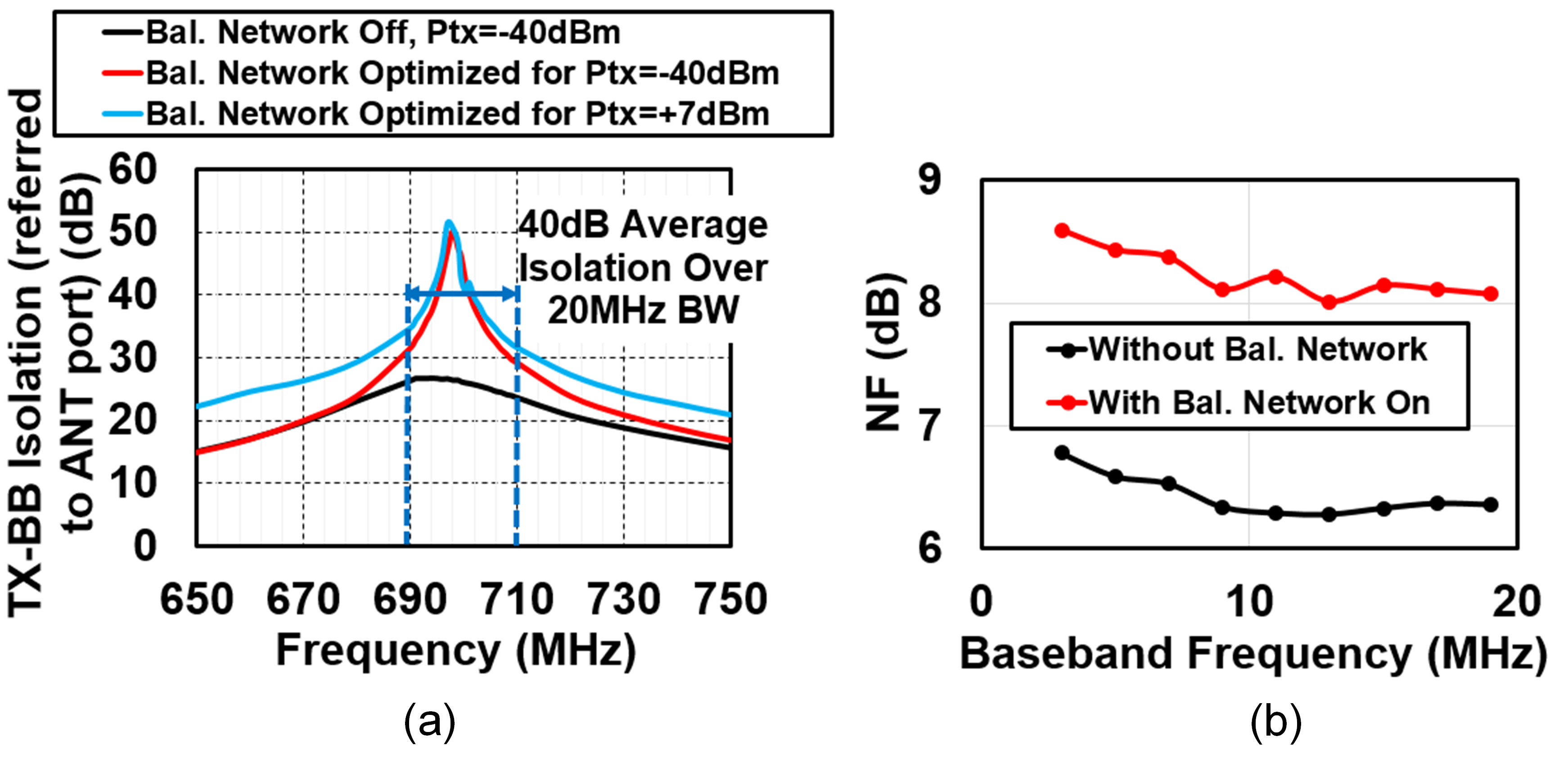}
	\caption{(a) Measured small and large signal TX-to-BB isolation after engaging the balance network. The balance network is reconfigured to maintain isolation for each TX power level. (b) Measured impact of the balance network on RX NF in the FD mode.}
	\label{fig:FD_meas}
\end{figure}

\begin{figure}[!h]
	\centering
	\includegraphics[keepaspectratio,width=0.65\linewidth]{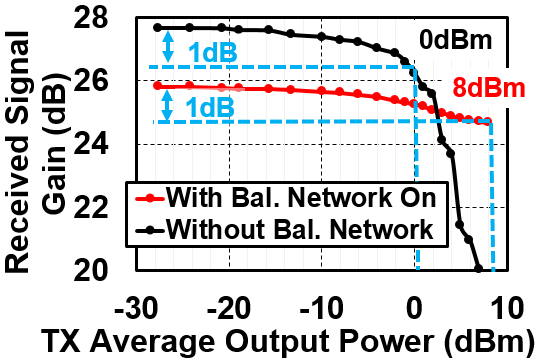}
	\caption{Measured ANT-to-RX-BB gain compression of a weak desired signal with and without balance network tuning versus varying TX output power level.}
	\label{fig:FD_meas_gain_comp}
\end{figure}

The block diagram and schematic of the 65 nm CMOS FD circulator-RX is shown in Fig. \ref{fig:chip_block_diagram}.

\subsection{Integrated Circulator}
The circulator was designed for tunable operation around 750~MHz in 65~nm CMOS technology. The $\frac{3\lambda}{4}$ line is miniaturized using three CLC sections implemented with on-chip MiM capacitors and off-chip air-core 8.9~nH inductors (0806SQ from Coilcraft, $Q_{L}$ \textgreater 100). The N-path filter uses 8-paths to increase the ANT-RX BB recombination gain and achieve harmonic cancellation for the 3rd and 5th harmonics. The capacitance of each path is 16~pF. Switch resistance for each of the sixteen transistors is around 3.5~$\Omega$. The sources/drains of switches are biased at 1.2~V and are DC coupled to the BB amplifiers (which run off a 2.4~V supply as is discussed later). The gates of the switches are AC coupled to the buffers and are biased at 1.35~V (DC level of a 12.5$\%$ pulse swinging from 1.2~V - 2.4~V). The balance network is designed using a parallel resistive bank (6 bits) and a parallel capacitive bank (5 bits). More complex balance networks as demonstrated in \cite{imec_TMTT_2016} can be used to increase the range of balancing that can be performed. An input clock at 4 times the operating frequency provides eight output phases in a Johnson-counter-based divide-by-4 block. Clock phase-shifting is performed for one set of switches prior to division using multiplexed digital delay cells with analog varactor-based fine-tuning to cover a range of $-76^\circ$ to $78^\circ$ around the nominal phase setting at 750~MHz based on schematic-level simulations at the typical corner. Simulations also reveal that in the worst-case (slow-slow corner) the clock path degrades the NF by about 0.25~dB.

\begin{figure}[!t]
	\centering
	\includegraphics[keepaspectratio,width=1\linewidth]{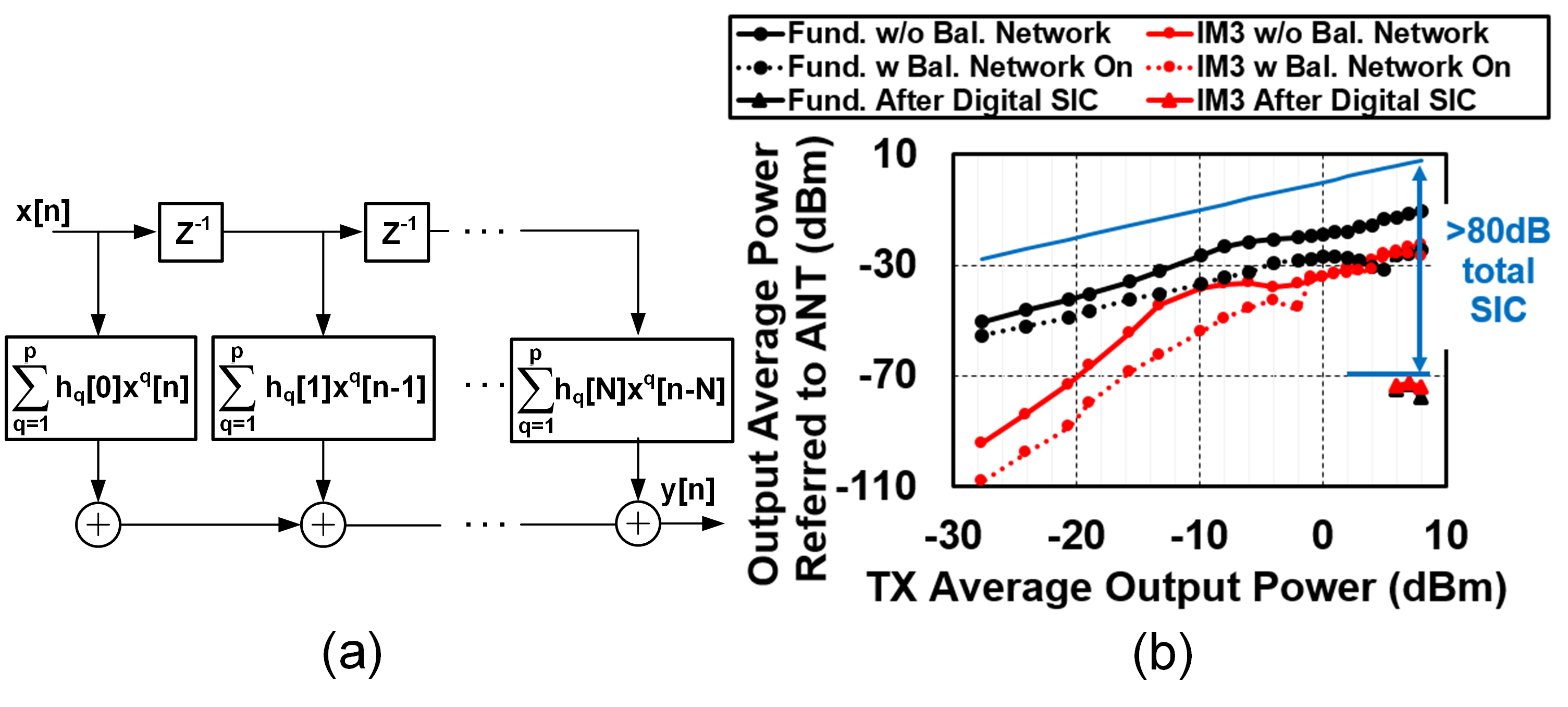}
	\caption{(a) Nonlinear tapped delay line used for digital SIC, and (b) measured two-tone TX test tracking the SI and its IM3 distortion at the RX BB output with SI suppression across antenna and digital domains.}
	\label{fig:FD_IIP3_digSIC}
\end{figure}

\begin{figure}[!h]
	\centering
	\includegraphics[keepaspectratio,width=\linewidth]{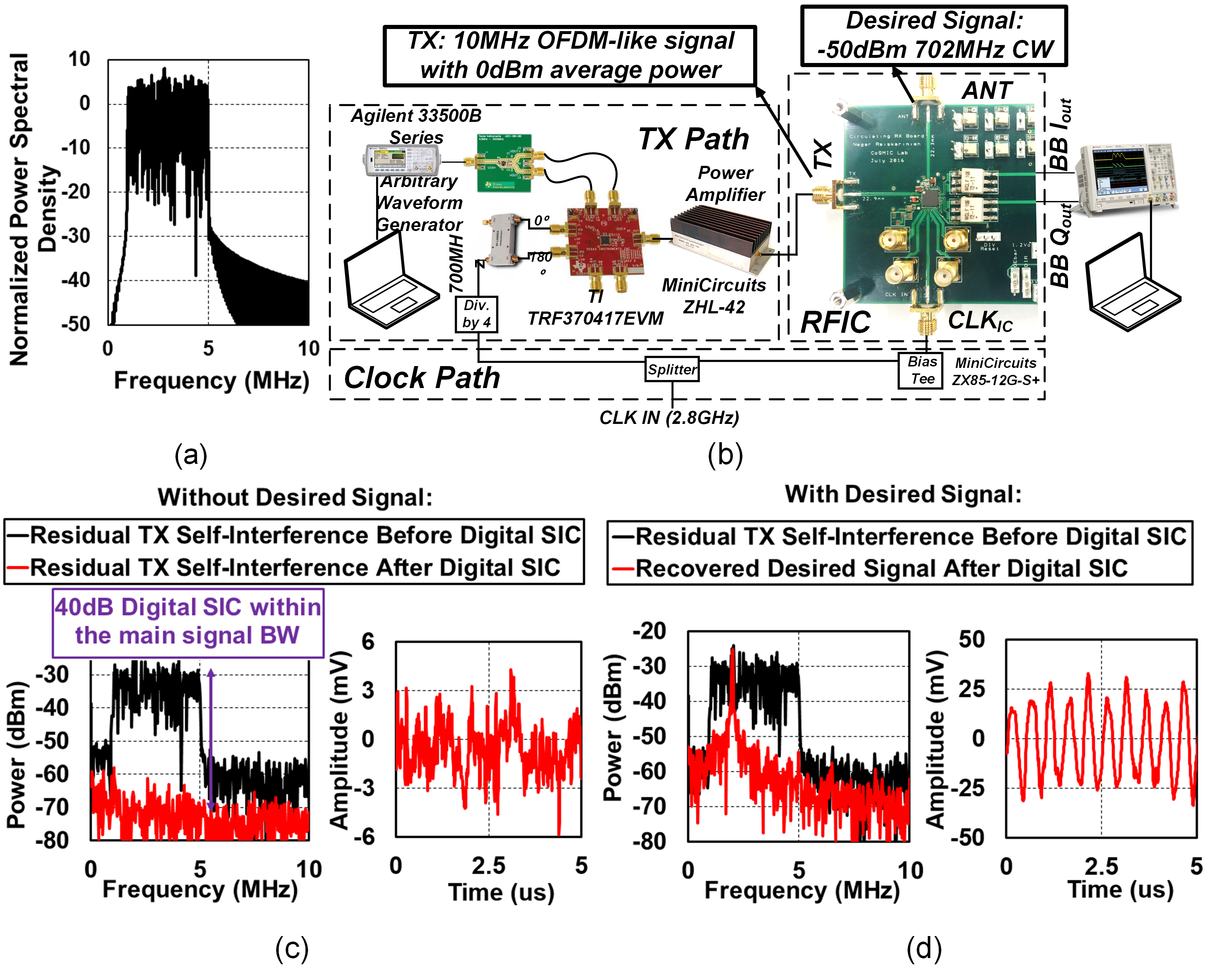}
	\caption{(a) Simulated BB frequency domain representation of the generated pulse-shaped OFDM-like signal used in the FD demo. (b) FD demonstration setup. FD demo results: a -50~dBm weak desired signal is received while transmitting a 0dBm average-power OFDM-like signal. Power spectral density and time-domain representation of the RX BB output before and after digital SIC are shown (c) without the desired ANT signal, and (d) with the desired ANT signal. The single tone ANT signal is recovered after digital SIC.}
	\label{fig:demo_setup}
\end{figure}
\subsection{Integrated Receiver}
The baseband circuitry consists of a baseband amplification stage and harmonic recombination circuitry. All the BB circuitry use thick-oxide devices and run off a 2.4~V supply to increase the power-handling of the circ.-RX. Four differential BB amplifiers are implemented, each using an inverter with large resistive feedback for self biasing and a common mode feedback circuit. A 5-bit variable resistor is added at the output to control the gain and BW. 

Since the circulator is based on an 8-path filter, the BB signals have to be recombined to provide differential I/Q outputs. The outputs of the BB amplifiers are connected to the harmonic-recombination variable-gain $G_m$ cells. The $G_m$ cells are implemented as open-drain differential pairs with switchable devices for 5-bit variable gain. Note that the variable-gain control is common for all the $G_m$ cells, and therefore, is not intended for harmonic rejection calibration. The differential outputs of the $G_m$ cells are connected to off-chip baluns for testing. The I+/- outputs are created by combining the differential 0/180$^\circ$ phases with a weight of 1, 45/225$^\circ$ phases with a weight of $\sqrt 2/2$ and 135/315$^\circ$ phases with a weight of $-\sqrt 2/2$. Similarly, the Q+/- outputs are generated by assigning a weight of 1 to the 90/270$^\circ$ phases, and $\sqrt 2/2$ weights to 45/225$^\circ$ and 135/315$^\circ$. This weighting cancels harmonics of the following orders: 3rd and 5th, 11th and 13th, and so on \cite{Molnar_JSSC2010}. Similar to prior work, the harmonic rejection is limited to the precision of the $\sqrt 2/2$  implementation and the mismatch between the devices. In this work, this scaling factor has been incorporated into the relative width of the NMOS devices in the $G_m$ cells. The overall harmonic rejection of the circulator-receiver is expected to be more, due to the band-limited nature of the circulator transmission lines implemented in this work.

\section{Measurement Results}

The chip microphotograph of the 65~nm CMOS circulator-receiver is shown in Fig.~\ref{fig:Chip_Photo}. It has an active area of 0.94~mm$^2$ and is mounted in a 40-pin QFN package. An SMD CLC-based fixed ANT tuning network as shown in Fig.~\ref{fig:matching_network} has been incorporated on the printed circuit board (PCB) performing a minor transformation (VSWR=1.2) to compensate for QFN parasitics and achieve a reasonable nominal TX-to-BB isolation when the balance network is off.

\subsection{Integrated Circulator}

The measured two-port TX-to-ANT S-parameters of the circulator for a clock frequency of 750~MHz are shown in Fig.~\ref{fig:circ_SP_IIP3}(a). Note that the RX is not available as a separate RF port, and hence the circulator's ANT-to-RX and TX-to-RX performance cannot be measured directly using S-parameters. These performance metrics of the circulator are a part of the circ.-RX measurements reported in the next subsection. The minimum measured TX-to-ANT loss is 1.8~dB with only 0.1~dB degradation in a 100~MHz BW around the center frequency. Both TX and ANT ports are well matched across a 300~MHz BW. Fig.~\ref{fig:circ_SP_IIP3}(b) shows the TX-to-ANT loss at the center frequency as clock frequency is tuned. For each clock frequency, phase-shift tuning between the clocks on either side is used to minimize the losses. The off-chip SMD-based matching network also improves the frequency range across which the $S_{21}$ remains below 3~dB by making the ANT impedance closer to the nominal $50~\Omega$. Less than 3~dB loss is maintained over 610-975~MHz by using the off-chip matching network. The measured circulator in-band (IB) TX-to-ANT IIP3 is +32.3~dBm as shown in Fig.~\ref{fig:circ_SP_IIP3}(c). This is higher compared to the implementation in \cite{NRK_JSSC2017} due to the lower $R_{sw}$ and the higher initial TX-to-RX isolation due to the SMD-based matching network. 

\begin{table*}[!t]
	\centering
	\caption{Performance summary and comparison with state-of-the-art full-duplex receivers.}
	\includegraphics[keepaspectratio,width=1\linewidth]{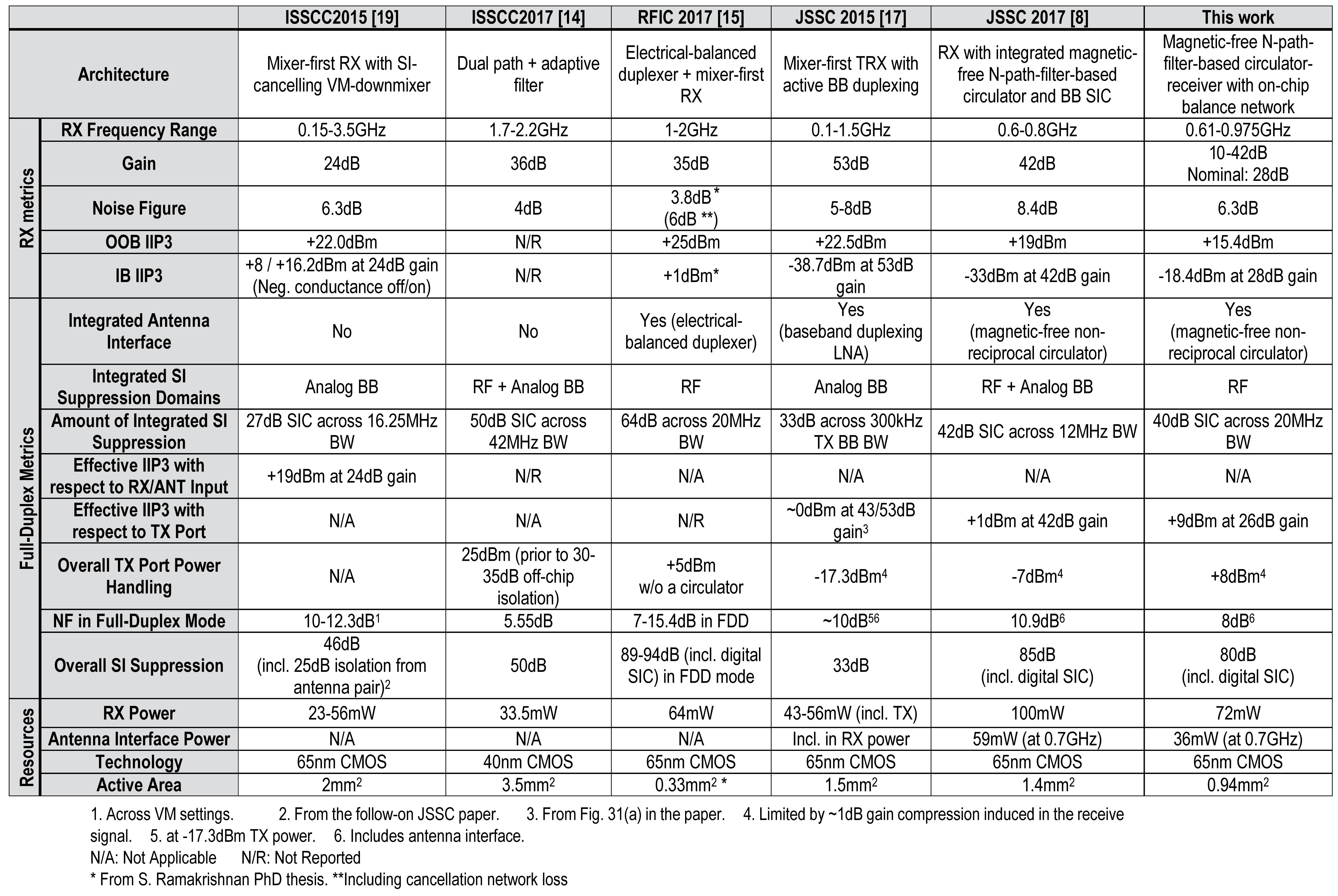}
	\label{tab:Comparison_table}
\end{table*}

\subsection{Circulator-Receiver Measurements without the Balance Network}
\vspace{-5pt}
The circulator-receiver operates over the frequency range of the on-chip integrated circulator, namely, 610-975~MHz, with a measured variable gain of 10-42~dB and a nominal gain of 28~dB. The RF BW of the circulator varies over 10-32~MHz as the gain is varied (Fig.~\ref{fig:RX_meas}(a)). The measured in-band IIP3 of the circ.-RX is -18.4~dBm at the nominal gain setting, and the measured out-of-band IIP3 is +15.4~dBm at a 500~MHz offset frequency as shown in Fig.~\ref{fig:RX_meas}(b). Our measured OOB IIP3 value is similar to that of mixer-first receivers \cite{Molnar_JSSC2010,CornellFullDuplex_JSSC15}, and can be improved to reach higher values with the use of more stages of low-pass-filtering through the receiver chain. The measured TX-to-BB isolation referred to the ANT port is better than 20~dB over more than 50~MHz BW ($6.7\%$ fractional BW) for a -40~dBm TX excitation after optimizing the on-board matching network. A receiver NF of 6.3dB is measured, and is comparable to or better than that of mixer-first receivers used for FD \cite{CornellFullDuplex_JSSC15,TwenteSICRX_JSSC2015} thanks to the relaxed receiver 50~$\Omega$ matching requirements and the effect of noise circulation. It should be emphasized that our reported NF encompasses both the receiver and an on-chip shared-ANT interface. 

\vspace{-10pt}
\subsection{Circulator-Receiver Measurements with the Balance Network}
\vspace{-5pt}
Engaging and optimizing the balance network dramatically improves the isolation for both small-signal and large-signal TX excitations (Fig.~\ref{fig:FD_meas}(a)). The average large-signal isolation for a TX power of +7~dBm improves to 40~dB over 20~MHz BW. At the optimized balance network setting, the NF degrades by 1.7~dB to 8~dB (Fig.~\ref{fig:FD_meas}(b)). Enabling the balance network also enhances the TX power for 1~dB gain compression of a weak desired signal from 0~dBm to +8~dBm as shown in Fig.~\ref{fig:FD_meas_gain_comp}. Fig.~\ref{fig:FD_IIP3_digSIC}(b) depicts a two-tone TX test, tracking the TX main SI and its IM3 distortion at the RX output. We have also implemented digital SIC in Matlab after capturing the baseband signals using an oscilloscope (a 12b quantizer). The digital SIC cancels not only the main SI but also the IM3 distortion generated from the SI (Fig.~\ref{fig:FD_IIP3_digSIC}(a)). The total SIC for the main TX tones and TX IM3 tones are 86~dB and 80~dB at +8~dBm average TX power, respectively. The effective noise floor after digital SIC is -73~dBm. As mentioned before, providing an additional 20~dB BB SIC, as shown before in \cite{NRK_JSSC2017}, would result in an overall noise floor of -93~dBm and enable a link range of 100~m at the operation frequency.

\vspace{-10pt}
\subsection{Comparison to the State of the Art}
\vspace{-5pt}

Table \ref{tab:Comparison_table} compares this paper to prior integrated FD RXs. This work has the highest TX power handling and isolation BW and the lowest NF among FD RXs with an integrated antenna interface. It also provides the benefit of embedding a balancing impedance to tune TX-to-RX isolation for minor antenna VSWRs. When compared with our prior work in \cite{NRK_JSSC2017}, this work has lower power consumption, better NF, higher tuning range, wider isolation BW, and higher effective IIP3 with respect to the TX port (and hence, higher TX power handling).
\section{FD Demonstration}
To demonstrate the effectiveness of the circulator-receiver architecture, a demonstration has been carried out in which a powerful modulated transmitted signal is canceled while a weak continuous-wave desired signal is received from the antenna port. 

An OFDM-like BB signal is generated at a sampling rate of 160~MSa/s, and it consists of 10 sub-carriers each with a bandwidth of 0.4~MHz occupying a total bandwidth of 5~MHz (DC-1~MHz has been omitted due to implementation limitations related to the high-pass cut-off frequency of the off-chip baseband baluns). The OFDM-like signal is pulse-shaped with square-root raised cosine (SRRC) filter with a roll-off factor of $\beta=0.22$. The total length of the OFDM-like signal is chosen to be 50000 samples with an extra 2000 samples to sync the received sequence to the transmitted signal. Fig.~\ref{fig:demo_setup}(a) shows a simulated frequency-domain representation of the generated pulse-shaped OFDM-like signal used in this demonstration. 

Fig.~\ref{fig:demo_setup}(b) shows the demonstration setup. The transmitter is built using off-the-shelf components as follows. An I/Q quadrature modulator (Texas Instruments TRF370417 EVM module) is followed by a Mini-Circuits ZHL-42 power amplifier. The BB OFDM-like signal is generated in MATLAB, and is fed to an Agilent 33500B arbitrary waveform generator (AWG), which is connected though a balun to the quadrature modulator. The clocks of the RFIC circulator-receiver and the transmitter are shared from a signal source running at 4$\times$ the frequency of operation (2.8~GHz) to lower the effects of uncorrelated phase-noise \cite{JZ_PN_TCAS2,Twente_PN,Rice_PN}. A separate frequency-division module is used to divide down the clock to 700~MHz for the transmitter.

First, to verify successful digital SIC of modulated signals, a 0~dBm average-power TX signal is applied to the TX port and the residual BB leakage is captured on the oscilloscope. In our digital SIC, nonlinear terms up to seventh order are considered, with a delay spread length of 45 samples, resulting in 315 total unknown canceller coefficients. An initial portion of the captured data (about $80 \mu sec$) is used to train the canceller coefficients. Fig.~\ref{fig:demo_setup}(c,d) shows the power spectral density at the RX BB output before and after the digital SIC of the TX leakage. 38~dB digital SIC and about 70~dB overall SIC has been achieved. Additionally, a -50~dBm weak continuous-wave desired signal at 702~MHz is applied to the ANT port, and is initially buried under the TX leakage at the RX BB before digital SIC. Once digital SIC is engaged, the desired signal is recovered, as can be seen in the time-domain and frequency-domain signal representations shown in Fig.~\ref{fig:demo_setup}(d).

\section{Conclusion} 
We have demonstrated a magnetic-free non-reciprocal N-path-filter-based circulator-receiver architecture for full-duplex wireless. The innovations include merging the commutation of the circulator with the down-conversion mixer to eliminate the receiver LNA and mixer, incorporation of an additional balancing impedance for fine tuning of the TX-to-RX isolation, and noise circulation of one set of commutating switches. A prototype 65~nm CMOS 750~MHz circulator-RX is presented which achieves 40~dB SIC across 20~MHz BW with lower NF and power consumption than prior works. In conjunction with digital SIC, the FD circulator-RX demonstrates 80~dB overall SI suppression for up to +8~dBm TX average output power. 

Future research directions include further improvements of the circulator's linearity performance and TX power handling and improved balancing schemes to account for a wider range of antenna impedance mismatches. At the system level, extending the research on integrated FD RXs to FD multi-antenna systems is of significant interest. 


\bibliographystyle{IEEEtran}
\bibliography{IEEEabrv,References}

\end{document}